\documentclass[a4paper,11pt]{article}

\usepackage{jheppub} 

\usepackage{lineno}
\usepackage{bbm}
\usepackage{amsmath,amssymb}
\usepackage{physics}
\usepackage{dsfont}
\usepackage{graphicx}
\usepackage{hyperref}
\usepackage[dvipsnames]{xcolor}
\usepackage{xstring}
\usepackage{forloop}
\usepackage{bm}
\usepackage{mathrsfs}
\usepackage{verbatim}
\usepackage{slashed}

\definecolor{tred}{cmyk}{0, 0.8, 0.62, 0.1}
\definecolor{tblue}{cmyk}{1, 0, 0.2, 0.1}
\definecolor{tgreen}{cmyk}{0.65, 0, 1.0, 0.1}

\def\beq{\begin{eqnarray}}
\def\eeq{\end{eqnarray}}
\def\bea{\begin{eqnarray}}
\def\eea{\end{eqnarray}}

\def\tev{\, {\rm TeV}}
\def\gev{\, {\rm GeV}}
\def\mev{\, {\rm MeV}}

\newcommand{\gsim}{\lower.7ex\hbox{$\;\stackrel{\textstyle>}{\sim}\;$}}
\newcommand{\lsim}{\lower.7ex\hbox{$\;\stackrel{\textstyle<}{\sim}\;$}}

\def\mpl{M_{\rm Pl}}

\newcommand{\nnmb}{\nonumber}
\newcommand{\del}{\partial}

\newcommand{\lrf}[2]{\left(\frac{#1}{#2}\right)}

\newcommand{\lag}{\mathscr{L}}

\newcommand{\btg}{\tilde{b}_>}
\newcommand{\btl}{\tilde{b}_<}
\newcommand{\btd}{\tilde{b}_\Delta}
\newcommand{\mvev}{m(\phi_0)}
\newcommand{\phivev}{{\phi_0}}
\newcommand{\nff}{N_\psi}
\newcommand{\fdm}{f_{\rm DM}}

\title{\boldmath On the Origins of Varying Gauge Couplings}

\author{Carlos Henrique de Lima,$^{a}$~}
\affiliation{$^{a}$TRIUMF, 4004 Wesbrook Mall, Vancouver, BC V6T 2A3, Canada}
\affiliation{$^{b}$School of Physics, Korea Institute for Advanced Study, 85 Hoegi-ro, Dongdaemun-gu, Seoul, 02455, Republic of Korea}

\emailAdd{cdelima@triumf.ca}

\author{David McKeen,$^{a}$~}
\emailAdd{mckeen@triumf.ca}

\author{David E. Morrissey,$^{a}$~}
\emailAdd{dmorri@triumf.ca}

\author{and Michael Shamma$^{a,b}$}
\emailAdd{mshamma@kias.re.kr}

\abstract{
Variations in the gauge couplings of the Standard Model have been searched for experimentally and proposed as solutions to open questions in fundamental physics. Varying gauge couplings can arise from a dynamical scalar field coupled to the gauge kinetic term. This mechanism has been invoked extensively, assuming a minimal linear coupling of the scalar to the gauge bosons. In this work, we investigate how this operator is generated from the ultraviolet perspective.  We argue that in weakly coupled, renormalizable completions in four spacetime dimensions, gauge invariance forces the leading dependence of the effective gauge coupling on the scalar to be logarithmic rather than linear. The gauge coupling evolution in these scenarios can be entirely described by the renormalization-group running with dynamical mass thresholds. Beyond leading order or four dimensions, we provide examples showing that more general behavior is possible. Finally, we discuss the phenomenological implications of dynamically evolving gauge couplings, particularly in early-universe settings.
}

\begin{document}

\maketitle
\flushbottom

\section{Introduction}
\label{sec:intro}

For nearly as long as it has been recognized that the laws of nature could be described in terms of a finite number of fundamental constants, scientists have asked whether these constants are, in fact, constant~\cite{Dirac:1979vf,Bekenstein:1982eu}. Quantum field theory~(QFT) provides a definitive answer to this question, where, at the cost of making finite predictions, the fundamental constants in QFT depend on the energy scale at which they are measured, as described by the Renormalization Group~(RG)~\cite{Gell-Mann:1954yli,Callan:1970yg,Wilson:1974mb,Wilson:1971bg,Wilson:1971dh,Curtright:2010hq}. For example, in four spacetime dimensions the leading running of the gauge couplings is logarithmic in energy, with the rate of variation determined by the number and charges of the fields that remain dynamical at that energy scale. This prediction of QFT has been verified experimentally for the strong~\cite{Huston:2023ofk}, electromagnetic~\cite{OPAL:2005xqs,L3:2005tsb}, and weak~\cite{Erler:2004nh} forces.

While this variation is remarkable, many authors have investigated the possibility of much larger temporal variations of the fundamental parameters over the history of the universe up to the present time, beyond what is predicted by the renormalization group evolution with a decreasing energy scale set by the cosmological temperature. Precision laboratory experiments have put very strong limits on parameter variations today~\cite{Olive:2002tz,Olive:2007aj,Uzan:2010pm,Uzan:2024ded}, while astrophysical and cosmological observations constrain variations from more recent times~\cite{Webb:1998cq,Murphy:2000pz,Webb:2000mn,Murphy:2003hw}, back to recombination~\cite{Hart:2019dxi,Seto:2023yal,Chluba:2023xqj}, and all the way to the era of primordial big bang nucleosynthesis~\cite{Campbell:1994bf,Bergstrom:1999wm,Coc:2006sx,Burns:2024ods}. Parameter variations in the early universe before BBN are much less constrained, and large dynamical changes in the SM gauge and Yukawa couplings have been proposed to enable dark matter creation~\cite{Berger:2020maa,Howard:2021ohe}, axion physics~\cite{Dvali:1995ce,Choi:1996fs,Heurtier:2021rko}, baryogenesis~\cite{Baldes:2016gaf,Ellis:2019flb,Croon:2019ugf,Croon:2022gwq,Huang:2023gse}, gravitational waves~\cite{Baldes:2016rqn}, and eras with early $SU(3)_c$ confinement~\cite{Patel:2013zla,Ipek:2018lhm,Lu:2022yuc} or strongly-coupled $SU(2)_L$~\cite{Abbott:1981re,Abbott:1981yg,Berger:2019yxb,Lohitsiri:2019wpq,Bhalla-Ladd:2025agq}. In view of this body of work, an important question is how such coupling variations could arise within consistent quantum field theories, both in terms of the size of the variation of interest and the required time to complete the transition. We investigate this question here with a primary focus on varying gauge couplings.

Many scenarios of dynamical gauge couplings have tied the variation to the vacuum expectation value~(VEV) of a singlet scalar field $\phi$. The leading effective operator of relevance consistent with gauge invariance is then
\beq \label{eq:dim5}
\lag \ \supset \ -\frac{1}{4}\left(\frac{1}{g^2}-c_g\frac{\phi}{\Lambda_g}\right)F^a_{\mu\nu}F^{a\,\mu\nu} \, ,
\eeq
where $F_{\mu \nu}^{a}$ is the field-strength tensor of the gauge boson, $\Lambda_g$ denotes the scale of new physics linking $\phi$ with the gauge group, and $c_g$ is the new operator's Wilson coefficient. If the scalar field VEV changes with respect to other parameters, the dimension-five operator above leads to an effective field-dependent gauge coupling
\begin{align}\label{eq:fdp}
   \frac{1}{g^{2}_{\text{eff}}(\mu,\phi)} = \frac{1}{g^{2}(\mu)} - c_{g} \frac{\phi}{\Lambda_g} \ .
\end{align} 
The form of Eq.~\eqref{eq:fdp} is invoked in many applications that rely on varying gauge couplings and suggests that large and rapid changes in $g_{\rm eff}$ are possible within the range of validity of the EFT expansion. 

A more complete perspective on the operator of Eq.~\eqref{eq:dim5} can be obtained by studying explicit ultraviolet~(UV) completions that give rise to the interaction. This is the approach we take in this work, and it leads to our primary conclusion that in weakly-coupled, renormalizable four-dimensional QFTs, gauge invariance and renormalization group structure force the leading dependence of the low-energy gauge coupling on a singlet scalar VEV to be logarithmic. A well-known example of this arises when the gauge theory with a singlet scalar is augmented by new states that are charged under the gauge group and whose masses depend on $\phi$~\cite{Dine:2002ir,Chacko:2002mf,Dent:2003dk,Chacko:2012sy,Gan:2023wnp,Davoudiasl:2018ltz}. Integrating out these charged states at their mass scale $m(\phi)$, one obtains a low-energy ($\mu < m(\phi)$) operator of the form
\beq
\lag \ \supset \ -\frac{1}{4}\left[\frac{1}{g^2(\mu,\phi_0)}-\tilde{c}_g\ln\!\left(\frac{m(\phi)}{m(\phi_0)}\right)\right]F_{\mu\nu}^aF^{a\,\mu\nu} \ ,
\label{eq:fdplog}
\eeq
where $\phi_0$ is a fixed reference value of the scalar background and $\tilde{c}_g$ a model-dependent constant. The key feature here is the logarithmic dependence on $\phi$ through the charged particle mass, rather than purely linear. This reduces to the form of Eq.~\eqref{eq:fdp} with $c_g/\Lambda_g=\tilde{c}_g m^\prime(\phi_0)/m(\phi_0)$ for small field excursions from the reference $\phi_0$, but varies much more slowly than linearly for larger excursions. 

Even though the logarithmic form of Eq.~\eqref{eq:fdplog} was obtained for a specific UV model, we propose that the result is more general and arises as the leading-order correction from any UV completion of Eq.~\eqref{eq:dim5} in four spacetime dimensions. Our argument is based on matching scale anomalies between the effective theories above and below the mass threshold in the limit that all masses are generated by the scalar background. Surpassing logarithmic dependence requires going beyond leading order or renormalizability.

Avoiding the logarithmic dependence requires more ingredients; a direct example can be found in scenarios with extra dimensions, where there is a link between the geometry of the extra dimensions and the effective gauge coupling in the four-dimensional low-energy effective theory after compactification. Specifically, for a gauge field propagating in $n$ flat extra dimensions, there is the general relation~\cite{Arkani-Hamed:1998jmv,Antoniadis:1998ig,Arkani-Hamed:1998sfv} 
\begin{align} \label{eq:changeR}
\frac{1}{g^2}  \sim \frac{\text{Vol}_n}{g_n^2} \, ,
\end{align}
where $g$ is the gauge coupling in the four-dimensional effective theory, $g_n$ is the coupling of the $4+n$ dimensional gauge theory, and $\text{Vol}_n$ is the volume of the extra dimensions. If this volume depends on the VEV of a scalar, such as a radion, changes in this VEV can induce power-law changes in the effective gauge coupling~\cite{Marciano:1983wy}. Quantum (loop) corrections further modify the relation of Eq.~\eqref{eq:changeR} but respect the general scaling with the volume~\cite{Dienes:1998vh,Dienes:1998vg}.

The goal of this work is to investigate UV completions that generate the operator of Eq.~\eqref{eq:dim5} to better understand how it can arise, its range of validity, and the extent to which the gauge coupling can be changed in the early universe. In Sec.~\ref{sec:4d}, we analyze the charged matter portal and establish the logarithmic form at one loop. Section~\ref{sec:4dextra} extends the argument to gauge remnant models and theories with symmetries acting on the scalar, and presents a general perturbative argument at the level of scale anomaly matching. In Sec.~\ref{sec:4dother} we compare gauge and Yukawa coupling variations and identify a two-loop back-reaction channel. Section~\ref{sec:xd} treats extra-dimensional realizations. Phenomenological implications are discussed in Sec.~\ref{sec:pheno}, and we conclude in Sec.~\ref{sec:conc}.

\section{A Dynamical Gauge Coupling from Charged Matter}
\label{sec:4d}

In this section, we investigate the simplest class of models with dynamical gauge couplings in four dimensions consisting of new charged matter that obtains some of its mass from a singlet scalar $\phi$. We present the mechanism and its connections to the standard RG running of the gauge coupling. We then show a toy example of the expected behavior of the gauge coupling in this class of UV completions.

\subsection{Charged Matter Portal}

A simple UV completion of the operator of Eq.~\eqref{eq:dim5} consists of new charged matter fields that obtain mass from a gauge-singlet field $\phi$~\cite{Dine:2002ir,Chacko:2002mf,Dent:2003dk,Chacko:2012sy,Gan:2023wnp}. As an explicit example, we consider a gauge theory with group $G$ and light charged matter, together with a massive vector-like fermion $\psi$ charged under $G$ and a gauge-singlet scalar $\phi$. The scalar is assumed to couple to the massive fermion via the interactions
\beq
-\lag \ \supset \ (M + y \phi) \bar{\psi}\psi \ .
\label{eq:yukawa}
\eeq
This term produces a heavy fermion mass equal to
\beq
m(\phi) = |M+y\phi| \ .
\eeq
As the background value (VEV) of $\phi$ changes, so does the fermion mass $m(\phi)$. Because the fermion mass sets the threshold at which it contributes to the beta function, a change in the scalar background shifts the scale at which this contribution turns on. This effectively moves the RG matching point, inducing a field-dependent shift in the low-energy gauge coupling.

To demonstrate the impact on the gauge coupling, we follow the methods of Refs.~\cite{Shifman:1979eb,Shifman:1978zn}. Recall that the leading term in the (one-loop) gauge effective action is
\beq
\lag_{\rm eff} \ \supset \ -\frac{1}{4g^2(\mu)} F_{\mu\nu}^{a}F^{a\, \mu\nu} \ ,
\label{eq:lgeff}
\eeq
where $\mu$ is the renormalization scale in a mass-independent scheme such as $\overline{\rm MS}$. The dependence of the gauge coupling on $\mu$ follows from the renormalization group, and has a well-known solution at leading order in the absence of mass thresholds:
\beq
\frac{1}{g^2(\mu)} = \frac{1}{g^2(\Lambda)}+ 2\tilde{b}\ln\!\lrf{\mu}{\Lambda} \ ,
\eeq
where $\Lambda > \mu$ is a UV reference scale and
\beq
\tilde{b} = \frac{b}{(4\pi)^2} = \frac{1}{(4\pi)^2}\!\left[\frac{11}{3}C_2(G)-\frac{2}{3}\sum_{f}S_2(r_f)-\frac{1}{3}\sum_sS_2(r_s)\right]  \ .
\eeq
Here, $C_2(G)$ is the quadratic Casimir invariant of the gauge group $G$, $S_2(r)$ is the trace invariant of the representation $r$ normalized to $1/2$ for the fundamental representation of $SU(N)$, and the sums run over all representations $r_f$ of chiral fermions and $r_s$ of complex scalars. Only light particles with masses $m < \mu$ should be included in the sums, and at mass thresholds the running coupling can be taken to be continuous at leading order~\cite{Weinberg:1980wa,Hall:1980kf}. A dependence of the effective gauge coupling on the background value of $\phi$ then arises when $\mu$ falls below a charged mass threshold controlled by the scalar~\cite{Shifman:1979eb}. 

Turning to our specific example of a charged fermion with mass $m(\phi) = |M+y\phi|$, we have different RG coefficients $\btg$ for $\mu > m(\phi)$ and $\btl$ for $\mu < m(\phi)$. Running and matching the gauge coupling produces
\beq \label{eq:logmatter1}
\frac{1}{g^2(\mu,\phi)} &=& \frac{1}{g^2(\Lambda)} 
+ 2\btg\ln\!\lrf{\mu}{\Lambda} + 2\btd\ln\!\lrf{\mu}{m(\phi)}\Theta(m(\phi)-\mu) \ ,
\eeq
where $\btd \equiv \btl - \btg$ is positive, and we have assumed that the UV gauge coupling at scale $\Lambda \gg m(\phi)$ is independent of $\phi$. The last term in this expression depends on the background value of $\phi$ as claimed. Including more charged states whose mass depends on $\phi$ generalizes by summing over their individual contributions.

It is helpful to rewrite Eq.~\eqref{eq:logmatter1} in terms of the gauge coupling that would be extracted from low-energy data in the universe today: $g(\mu_0,\phi_0)$ defined at some low reference scale $\mu_0$ with the value of the scalar equal to its current value, $\phi \to \phi_0$ such that $m(\phi) \to m(\phi_0)$. The field-dependent gauge coupling can be written with respect to this reference as
\begin{align} \label{eq:logmatter2}
 \frac{1}{g^2(\mu,\phi)} &= \frac{1}{g^2(\mu_0,\phi_0)} + 2\btg\ln\!\lrf{\mu}{\mu_0} \\
 & + 2\btd\left[\ln\!\lrf{\mu}{m(\phi)}\Theta(m(\phi)-\mu)
-\ln\!\lrf{\mu_0}{m(\phi_0)}\Theta(m(\phi_0)-\mu_0)\right] \, .
\nnmb
\end{align}
We will always assume a reference scale $\mu_0 < m(\phi_0)$. This implies
\begin{align}
 \frac{1}{g^2(\mu,\phi)} &=    \frac{1}{g^2(\mu_0,\phi_0)}
+ 2\btl\ln\!\lrf{\mu}{\mu_0} -2\btd\ln\!
\lrf{\max\{m(\phi),\mu\}}{m(\phi_0)} \ .
\end{align}
The first two terms here combine to give the extrapolation of the low-scale gauge coupling measured at $\mu_0$ to the scale $\mu$ with fixed background $\phi_0$ in the absence of new matter. All the dependence on $\phi$ comes from the last term. We see that the gauge coupling depends on $\phi$ only when $\mu < m(\phi)$, and that the argument of the mass-dependent logarithm in Eq.~\eqref{eq:logmatter2} never becomes singular. As for Eq.~\eqref{eq:logmatter1}, if there are multiple massive charged fields, their individual contributions should be summed over.

Combining Eq.~\eqref{eq:logmatter2} with Eq.~\eqref{eq:lgeff}, the field-dependent correction obtained here has precisely the form of Eq.~\eqref{eq:fdplog} when $\mu < m(\phi)$. For small fractional mass variations $|\Delta m(\phi)|/\mvev \equiv |m(\phi)-m(\phi_0)|/\mvev \ll 1$, the logarithm can be reliably expanded to produce an effective gauge action of the form of Eq.~\eqref{eq:dim5}:
\beq\label{eq:ad}
\lag_{\rm eff} \ \supset \ -\frac{1}{4}\left[\frac{1}{g^2(\mu,\phi_0)}
-2\btd
\!\left.\frac{1}{m}\frac{dm}{d\phi}\right|_0(\phi-\phi_0)
+ \ldots\right] F_{\mu\nu}^{a}F^{a \, \mu\nu} \ ,
\eeq
Using Eq.~\eqref{eq:ad}, we can identify the coefficient $c_g$ and the mass scale $\Lambda_g$ of the effective operator of Eq.~\eqref{eq:dim5} within our specific UV completion. However, the full theory allows us to resum a tower of EFT operators to obtain a logarithm that is valid for large variations $|\Delta m|/m(\phi_0)\gtrsim 1$ as well. Indeed, the physical origin of dynamical gauge couplings in this theory can be understood entirely in terms of the impact of field variations on the RG evolution of the gauge coupling.

\subsection{Dynamical Gauge Couplings in the Early Universe}

While the scale dependence of the gauge couplings at fixed background field values is accounted for by standard RG evolution, changes in the field value of $\phi$ modify charged matter thresholds and can induce a further variation in the gauge couplings.  In this section, we present a toy example that illustrates this dependence in such scenarios. Our goal is not to build a realistic cosmological model, but rather to demonstrate potential evolution pathways of gauge couplings as the universe expands and cools. We focus on the charged matter model presented above with a $U(1)$ gauge theory and a real scalar $\phi$ together with $\nff$ unit-charged fermions with equal masses $m(\phi)= |M+ y\phi|$.

The gauge coupling in this theory can be viewed as a function of both $\mu$ and $\phi$, as described by Eq.~\eqref{eq:logmatter2}. However, in the hot early universe, the relevant RG scale $\mu$ and scalar expectation value $\phi$ are both related to the temperature. Specifically, perturbation theory in the thermal background is optimized for $\mu \sim T$~\cite{Dolan:1973qd,Weinberg:1974hy,Kneur:2015uha,Blaizot:2014ffa,Matsumoto:1983jq}. Furthermore, the expectation value of the scalar field is determined by the thermal effective potential, which itself depends on temperature, so that $\phi = \phi(T)$. Together, this implies a thermal field trajectory in the two-dimensional $(\mu,\,\phi)$ space on which we should evaluate $g(\mu,\phi)$. 

The overall dynamics depend on the specific $\phi$ trajectory; in this example, we assume a representative thermal potential for $\phi$ of the form
\beq
V_\phi(\phi,T) = -\frac{1}{2}\mu_\phi^2\phi^2 +\frac{\lambda}{4}\phi^4 + \frac{1}{2}\xi\,T^2\phi^2 \ .\label{eq:thermalpot}
\eeq
The last term is characteristic of the leading (model-dependent) thermal correction from light matter coupled to $\phi$~\cite{Anderson:1991zb}. With this potential, the expectation value $\phi(T)$ is 
\beq\label{eq:profile}
\phi(T) = \phi_0\,\sqrt{1-\lrf{T}{T_c}^{\!2}}\;\Theta(T_c-T) \ ,
\eeq
where 
\beq
\phi_0 = \sqrt{\frac{\mu_\phi^2}{\lambda}} \ ,
\qquad
T_c = \sqrt{\frac{\mu_\phi^2}{\xi}} = \sqrt{\frac{\lambda}{\xi}}\,\phi_0 \ .
\eeq
The potential yields a second-order phase transition with critical temperature $T_c$ such that $\phi(T\geq T_c) = 0$ evolves to $\phi(T)> 0 \to \phi_0$ as $T\to 0$. As the scalar background changes with temperature, so too does the fermion mass $m(\phi)$ and therefore the gauge coupling as described by Eq.~\eqref{eq:logmatter2}.

\begin{figure}[t!]
    \centering
 \includegraphics[width=0.65\textwidth]{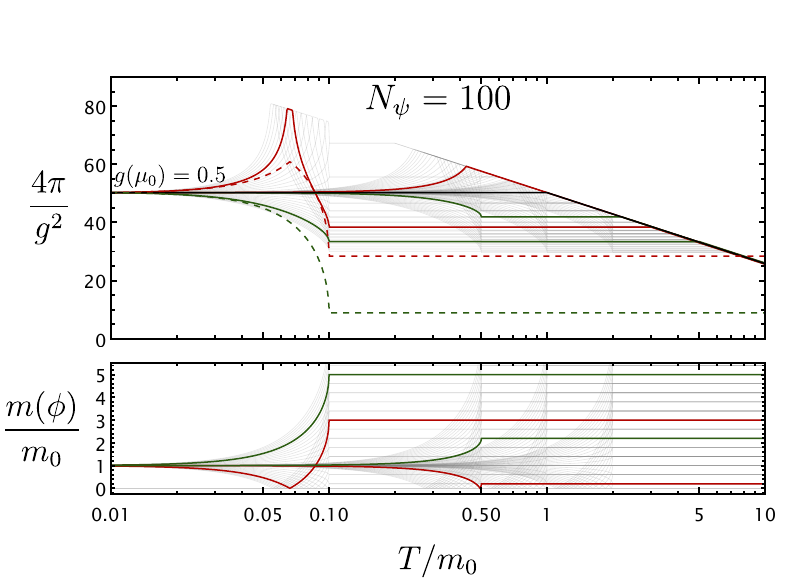}
    \caption{Inverse gauge coupling $4\pi/g^2$ (top panel) and fermion mass $m(\phi)$ (bottom panel) as functions of temperature $T$, normalized to the current $T=0$ mass $m_0\equiv m(\phi_0)$, for a toy Abelian model with $\nff=100$ vector-like fermions. Each curve corresponds to a different choice of $y\phi_0/m_0$ and critical temperature $T_c$, leading to distinct scalar field evolutions and hence different mass thresholds. The dashed lines show the naive linear expectation for the two pairs of benchmarks discussed in the text, while the solid red and green lines show the corresponding evolution in the full theory for specific parameter values.}
    \label{fig:p1}
\end{figure}

To make the example explicit, consider first the perspective of a zero-temperature observer who has measured the low-scale gauge coupling $g(\mu_0,\phi_0)$ and the fermion mass $m_0 \equiv m(\phi_0)$ in the vacuum $\phi=\phi_0$ at $T \simeq 0$ today at some low scale $\mu_0\ll m_0$. Depending on the scalar dynamics in the early universe, the gauge coupling at intermediate temperatures can be drastically different from the extrapolation of zero-temperature observables. We show this in the upper panel of Fig.~\ref{fig:p1}, where we plot $4\pi/g^2$ as a function of temperature normalized by the fermion mass today, $T/m_0$, for $g(\mu_0,\phi_0) = 0.5$, with $\nff = 100$ fermions to amplify the gauge change. The lower panel gives the corresponding evolution of the fermion mass. We show families of trajectories for four different critical temperatures $T_c/m_0 = [0.1,0.5,1,2]$ and a range of stepped values of $y\phi_0/m_0$ while taking $M > 0$ without loss of generality. We highlight two pairs of trajectories in particular. The first pair has $T_c/m_0= 0.1$ and $|y\phi_0|/m_0 = 4$, while the second pair has $T_c/m_0=0.5$ and $|y\phi_0|/m_0 = 1.2$. For both pairs, the green~(red) lines have $y\phi_0/m_0 > 0$~($<0$). The dashed lines show the evolution for these parameters in the linear approximation, and demonstrate that in these cases this approximation deviates significantly from the actual variation.

By construction, all the trajectories in Fig.~\ref{fig:p1} approach the same value at low temperatures. However, we also see that they align to the same curve once $T > T_c$. While each trajectory represents a different theory, the evolution with scale for $\phi = \phi_0$ is determined once $g(\mu_0,\phi_0)$ and $m_0$ are specified. Since all the theories have the same evolution equation for $\mu > m(\phi)$, fixing the IR coupling and normalizing the scale by $m(\phi_0)$ implies they also align in the UV as functions of $\mu/m_0$. Between these boundaries, the change in the gauge couplings can be understood as a movement from the RG trajectory for one value of the charged fermion mass to the trajectory of another.

There are some interesting features that we can learn from Fig.~\ref{fig:p1} that limit the application of this scenario for non-perturbative transitions. The subtlety is that the low-temperature value of the coupling on any given trajectory was already attained at a higher temperature earlier in the evolution. If a trajectory drives the coupling into the strongly coupled regime at some intermediate temperature, then the same trajectory was strongly coupled at all higher temperatures as well, and the calculation relies on extrapolating from a regime where perturbation theory has already broken down. Such transitions are not forbidden, but they lie outside the calculable regime. Large variations that remain within the weakly coupled regime throughout are fully under control and can still have non-trivial consequences.

\begin{figure}[t!]
    \centering
 \includegraphics[width=0.65\textwidth]{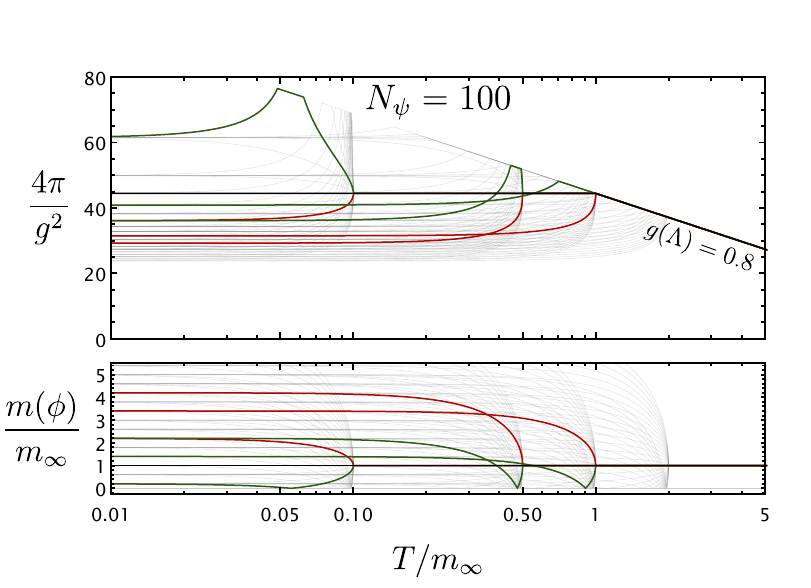}
    \caption{Same as Fig.~\ref{fig:p1}, but with the UV gauge coupling fixed at $g(\Lambda) = 0.8$ and all dimensionful scales normalized to the UV fermion mass $m_\infty \equiv m(\phi=0)$ instead of IR quantities. }
    \label{fig:p2}
\end{figure}

A perspective that can be more useful to see this limitation comes from the viewpoint of a UV observer. In the UV, the measured gauge coupling is unique and defines the gauge theory. In this case, the family of theories with different masses flows to different values at low energy. We construct such a setup in Fig.~\ref{fig:p2}, with the same mass profiles, but now normalized to the high-energy mass of the fermion, $m_{\infty}\equiv M$. In this way, it becomes more explicit how the low-energy gauge coupling is dependent on everything that occurred as the theory flows from the UV to the IR.

From these toy examples, we can see that the linear expectation does not properly describe these scenarios, which can be understood by the breakdown of the EFT from the new heavy states being close to where the dynamics are happening. Performing the analysis with the dynamical RG picture shows that a plethora of interesting dynamics can occur, but with stronger limitations than those derived from the EFT picture. The biggest limitation is the connection between the mass scale that is changing and the energy scales being probed. If the effect of the degrees of freedom is already fully accounted for, i.e., they are changing, but at energy scales much above their scale, there is no residual effect at low energies. This brings constraints on model building for the Standard Model cases, which we further explore in Sec.~\ref{sec:pheno}.

In the scenarios that we show here, the gauge coupling becomes strongly coupled at high energies. This is just a consequence of the beta function of this model, where the additional fermions in the UV drive the theory to a Landau pole. This is not necessarily universal, and we show in the next section how to construct theories where the slope changes differently.

\section{Other Realizations in Four Dimensions} \label{sec:4dextra}

The charged matter model presented above demonstrates how variations in a singlet field can induce changes in the gauge coupling. In this section, we consider other realizations of dynamical gauge couplings in renormalizable four-dimensional theories. We argue that the mechanisms relevant at leading perturbative order fall into a small number of classes. We first consider scenarios where the gauge structure changes across a threshold due to Higgsing to a smaller subgroup. Next, we relate the previous charged matter model to the dependence of the SM gauge couplings on the Higgs VEV. We then examine the impact of symmetries acting on the scalar field inducing the dynamical gauge variation. Finally, we present a general argument that a logarithmic dependence on the varying field is a generic feature of weakly-coupled renormalizable theories at leading order.

\subsection{Gauge Remnant Model\label{sec:remnant}}

An implicit assumption in the charged matter model we considered above was that the underlying gauge structure remains unchanged through the massive threshold. Here, we relax this assumption and study a scenario where a massive threshold changes the gauge structure, reducing it to a non-trivial subgroup of lower rank. Despite this difference, we show that the variation of the remnant low-energy gauge coupling has the same logarithmic dependence on the threshold as the charged matter model.

Consider a gauge theory with group $SU(N+1)$ and $N\geq 2$ that is spontaneously broken to the subgroup $SU(N)$ by the VEV of a fundamental scalar $\Phi$ with magnitude $\phi$. Of the $2N+2$ real degrees of freedom of $\Phi$, $2N+1$ of them are eaten by the massive vector bosons. The single remaining component is a real scalar that is a singlet under the remnant $SU(N)$ subgroup. The Higgsed vector bosons have masses that scale as $m(\phi) \sim g\phi$, and for simplicity we assume that the real scalar mass is similar.

The low-energy $SU(N)$ gauge coupling depends on the symmetry-breaking-induced mass scale $m(\phi)$. As before, its value can be obtained by running and matching the gauge couplings at the massive threshold. At scales $\mu \gg m(\phi)$ the full $SU(N+1)$ is manifest, while for $\mu \ll m(\phi)$ the massive vectors and singlet scalar should be integrated out and the explicit gauge structure is $SU(N)$. The leading-order matching condition for the gauge couplings at the threshold $m(\phi)$ is $g_{N\!+\!1}^2(m)=g_N^2(m)\equiv g^2(m)$. With this boundary condition, the expression for $g^2(\mu,\phi)$ for $\mu < m(\phi)$ in the $SU(N)$ effective theory is given precisely by Eq.~\eqref{eq:logmatter1}. This again implies a logarithmic coupling of $\phi$ to the gauge field strength in the low-energy effective action.

To evaluate $g^2(\mu,\phi)$ in the low-energy theory, we need the RG coefficients $\btg$ and $\btl$ above and below the threshold. They are 
\beq
\btg &=& \frac{1}{(4\pi)^2}\bigg[\frac{11}{3}(N+1) - \frac{1}{6}\bigg]+\btg^{\rm matt} \, , \\
\btl &=& \frac{1}{(4\pi)^2}\bigg[\frac{11}{3}N\bigg] +\btl^{\rm matt}\nnmb \, ,
\eeq
where $\tilde{b}_{>,<}^{\rm matt}$ refer to the contributions from matter fields charged under the $SU(N+1)$ and $SU(N)$ groups. If all the matter representations are fundamentals under $SU(N+1)$ and do not obtain mass from $\Phi$ then $\btd^{\rm matt}=\btl^{\rm matt}-\btg^{\rm matt} =0$; otherwise $\btd^{\rm matt} > 0$. Assembling these pieces, the effective gauge coupling is
\beq
\frac{1}{g^2(\mu,\phi)} &=& \frac{1}{g^2(\Lambda)} + 2\btg\ln\!\lrf{\mu}{\Lambda}
\label{eq:logremnant}
\\
&&\hspace{-0.5cm}+ \left(-\frac{22}{3}+2\btd^{\rm matt}\right)\ln\!\lrf{\mu}{m(\phi)}\Theta(m(\phi)-\mu) \ .
\nnmb
\eeq
The coefficient of the $\phi$-dependent logarithmic term here is $2\btd = 2(\btl - \btg)= -7+2\btd^{\rm matt}$. In contrast to our previous example with massive charged matter, where $\btd \geq 0$, the coefficient $\btd$ here can be positive or negative depending on the matter content. This is only possible thanks to the reduction in gauge rank through the mass threshold~\cite{Batra:2003nj}.

\subsection{Higgs Portal Model}

The SM itself realizes a form of the charged matter model by way of the Higgs field. Variations in the Higgs VEV lead to changes in the SM particle masses at tree-level. Since some of these particles are charged under the remnant electromagnetic and strong gauge groups after electroweak symmetry breaking, variations in the Higgs VEV produce changes in these gauge couplings at one-loop order. This property was used in Ref.~\cite{Shifman:1978zn,Shifman:1979eb} to obtain the leading low-energy effective couplings of the Higgs boson to photons and gluons.

The charged matter portal realized by the SM can be connected to a new singlet scalar, as proposed in Ref.~\cite{Piazza:2010ye}. This work introduces a real singlet $\phi$ with the coupling
\beq
-\lag \ \supset \ A\,\phi\,H^\dagger H \ ,
\eeq
where $A$ has a mass dimension of one. After electroweak symmetry breaking, the Higgs VEV generates an expectation value for $\phi$ and produces mass mixing between the physical excitation from $\phi$ and the would-be Higgs boson. This generates a coupling between the physical excitation from $\phi$ and photons and gluons. However, in this realization, the electromagnetic and strong couplings do not vary with the singlet VEV except possibly through its effect on the Higgs VEV. We also note that if there were no electroweak symmetry breaking, such as with a positive quadratic term in the Higgs potential, a dependence of the effective $SU(2)_L$ and $U(1)_Y$ gauge couplings on $\phi$ would be generated at scales below the mass of the charged Higgs doublet fields.

\subsection{Implications of Symmetries on the Low-Energy EFT} \label{sec:symmetry}

In the charged matter model considered in Sec.~\ref{sec:4d}, the scalar $\phi$ was not associated with any explicit symmetry. When such a scalar, whose VEV controls the masses of gauge-charged matter, transforms non-trivially under a symmetry, the specific dependence of the effective gauge coupling on the field value can be modified. We illustrate this with a specific example.

Consider a theory with a gauge-singlet complex scalar $\phi$ and a set of $\nff$ Dirac fermions $\psi_n$, $n=0,\ldots, \nff-1$ that transform under a complex representation of the gauge group. We take the theory to be invariant under a $\mathbb{Z}_{\nff}$ symmetry under which $\phi$ has charge 1 and $\psi_n$ has charge $n$, with all charges taken modulo $\nff$. The most general parity-conserving mass and Yukawa terms are then
\beq
-\lag \supset \sum_{n=0}^{\nff-1}M_n\bar{\psi}_n\psi_n + \left(\sum_{n=0}^{\nff-2}y_n\phi\bar{\psi}_{n+1}\psi_n + y_{\scriptscriptstyle  \nff-1}\phi\bar{\psi}_{0}\psi_{\scriptscriptstyle \nff-1}  + {\rm h.c.}\right) \ .
\label{eq:symmlag}
\eeq
When the scalar background $\phi$ is non-zero, the fermions mix through the mass terms. For scales $\mu$ less than all the fermion mass eigenvalues, Eq.~\eqref{eq:logmatter1} for the running gauge coupling generalizes to
\beq
\frac{1}{g^2(\mu,\phi)}= \frac{1}{g^2(\Lambda)}
+ 2\btg\ln\!\lrf{\mu}{\Lambda} + 2\btd\ln\!\lrf{\mu^{\nff}}{|\det m(\phi)|} \, ,
\label{eq:logshamma}
\eeq
where $\det m(\phi)$ is a shorthand for the product of mass eigenvalues with $m_i(\phi) > \mu$ and $\btd$ is the change in the beta-function coefficient from a \emph{single} fermion species. The underlying $\mathbb{Z}_{\nff}$ symmetry is encoded in the $\phi$ dependence of $m(\phi)$.

To see this feature explicitly, let us specialize to $\nff=3$. We obtain
\beq
\det m(\phi) = M_0M_1M_2 -M_0|y_1\phi|^2 - M_1|y_2\phi|^2 - M_2|y_0\phi|^2 + (y_0y_1y_2\phi^3+{\rm h.c.}) \ .
\eeq
All the terms here are consistent with the underlying $\mathbb{Z}_3$ symmetry. Furthermore, the $\phi^3$ term is proportional to the product of all the Yukawas, which is necessary since, as any $y_n\to 0$, the underlying Lagrangian respects a larger global $U(1)$ symmetry. Generalizing to other values of $\nff$, the mass product $m(\phi)$ has multiple terms with powers of $|\phi|^2$ together with a $\phi^{\nff}$ term proportional to the product of all the Yukawas, 
\beq
\det m(\phi)=\sum_{n=0,1,\dots}^{\lfloor \nff/2 \rfloor}a_n \left(\left|\phi\right|^2\right)^n + (-1)^{\nff-1} \left(\prod_{n=0}^{\nff-1}y_n\phi^{\nff}+{\rm h.c.}\right)\, ,
\eeq
where the $a_n$ coefficients are functions of $M_n$ and $y_n$ with, e.g., $a_0=\prod_n M_n$. Importantly, no term linear in $\phi$ appears in $\det m(\phi)$.

The implications of the scalar $\phi$ being connected to a symmetry can be seen by comparing the models above to a minimal charged matter theory with $\nff$ charged fermions where $\phi$ has no specific symmetry. The general expression of Eq.~\eqref{eq:logshamma} applies to both models. For large scalar field values, such that $|y\phi| \gg |M|$, both models have the same parametric dependence on $\phi$, going like $-2\nff\btd\ln(|\phi|/\mu)$. For smaller scalar field values, consider expanding the $\phi$-dependent term in fluctuations in the real direction around a real background VEV $\phi_0$, in the form $\phi = \phi_0+\Delta\phi$. We obtain 
\begin{align}
&\ln|\det m(\phi)| = \ln|\det m(\phi_0)|\\
&+ \Re\left\{\Tr\left(m^{-1}\frac{dm}{d\phi}\right)_{\!\!\phi_0}\!\right\} (\Delta\phi)
+\frac{1}{2}\Re\left\{
\Tr\left[m^{-1}\frac{d^2m}{d\phi^2}-\left(m^{-1}\frac{dm}{d\phi}\right)^2\right]_{\!\phi_0}
\right\}(\Delta \phi)^2
+ \ldots \, .
\nnmb
\end{align}
For generic theories where $\phi=\phi_0$ is not a point of enhanced symmetry, the trace of $(m^{-1}dm/d\phi)_{\phi_0}$ is non-zero, and the leading dependence of the gauge coupling on $\Delta\phi$ is linear. However, when the specific value of $\phi_0$ leads to the explicit realization of a symmetry, the first derivative $(dm/d\phi)_{\phi_0}$ must vanish, which can be seen easily for the $\mathbb{Z}_{\nff}$ model above when expanding around $\phi_0=0$. In this case, the leading dependence on $\Delta\phi$ is quadratic or higher, corresponding to operators of dimension-six or above in the low-energy EFT.

\subsection{General Arguments for Renormalizable Theories\label{sec:generalscale}}

A key feature in all the scenarios we have studied so far is a logarithmic dependence of the low-energy gauge coupling on the background field value $\phi$. We argue here that this is a generic property of renormalizable UV completions in four dimensions at leading perturbative order.

Gauge invariance severely limits the couplings of gauge bosons. Specifically, it forbids mass or kinetic mixing of gauge bosons with other fields in the absence of symmetry breaking (with the specific exception of kinetic mixing of Abelian vectors). In turn, this prevents changes to effective gauge couplings at tree level from mixing. A further implication of gauge invariance is that, up to operators of dimension four, it only allows gauge bosons to couple to charged matter. This matter feeds back on the gauge boson at one-loop order and contributes to the renormalization of the gauge coupling in relation to its gauge charge.

To alter the gauge coupling beyond its usual running, we should therefore look to change the properties of charged matter. Gauge invariance prevents changes to charges of matter fields and mixing among states with different charges, but can be consistent with variations in masses and non-gauge couplings. At one-loop order, only the masses impact the gauge running, and this has been the basis of the realizations we have studied so far.

Concentrating on mass variations of charged matter, we can make a more precise statement based on scale invariance. Suppose we have gauge theories based on groups $G_>\supseteq G_<$ that are matched at a massive threshold $m$. Let us assume further that the threshold arises from classically scale-invariant interactions together with the spontaneous breaking of scale invariance parametrized by $\phi$ (which may be fundamental or composite). To be precise, we take $\phi \to e^\lambda\phi$ under scale transformations together with the usual  $x\to e^{-\lambda}x$, $F_{\mu\nu}\to e^{2\lambda}F_{\mu\nu}$, and $\mu \to e^\lambda\mu$. Above the mass threshold, we assume a renormalizable theory with no dependence on $\phi$. The gauge portion of the scale anomaly coefficient obtained from the leading one-loop effective Lagrangian is then
\beq
\del_\mu D^\mu = \btg F_{\mu\nu}^a F^{a \, \mu\nu} \ ,
\eeq
where $D^\mu$ is the scale current. Below the mass threshold $m(\phi)$, the gauge part of the effective Lagrangian expanded in powers of the coupling has the form
\beq
\lag_{\rm eff} \supset -\frac{1}{4}\!\left[\frac{1}{g^2(\mu)}+ f(\phi)+\mathcal{O}(g^2)\right]\! F_{\mu\nu}^a F^{a \, \mu\nu}~~~~~
\eeq
for some function $f(\phi)$. This gives a leading contribution to the scale anomaly below the mass threshold of
\beq
\del_\mu D^\mu = \left[\btl - \frac{1}{2}\frac{df}{d(\ln\phi)}\right]\!F_{\mu\nu}^a F^{a \, \mu\nu} \ .
\eeq
Since we assume that classical scale invariance is only broken spontaneously, we expect a matching between the anomaly coefficients of light fields across the threshold. Applying this condition implies the relation
\beq
\frac{df}{d(\ln\phi)} = -2\big(\btl-\btg\big) = -2\btd \ .
\eeq
Solving, we find
\beq
f(\phi) = {\rm const.} -2\btd\ln\phi \ .
\eeq
This reproduces the logarithmic dependence found previously when $m(\phi) \propto \phi$. These arguments can be generalized to the case with explicit mass terms by treating them as spurions.

Before moving on, let us make two comments about these arguments. First, implicit in this discussion is that the kinetic term of the $\phi$ field is canonical (up to an overall normalization). This is needed for the scale transformation assumed for $\phi \to e^\lambda\phi$ to leave its kinetic term invariant, as well as for renormalizability. While changing field variables changes the logarithm of $\phi$ above into a different (and possibly non-logarithmic) dependence on the new field variable, it would come at the cost of a non-canonical kinetic term that obscures the renormalizability of the theory and implies a different scale transformation of the field variable. Equivalently, we find a logarithmic dependence on the invariant field distance with respect to the scalar metric defined through the kinetic term~\cite{Alonso:2015fsp,Helset:2020yio,Cohen:2020xca}. Our second comment is that our arguments here are strictly perturbative and only extend to one-loop (or leading log) order. At higher orders, the effective gauge coupling can depend on other quantities associated with charged matter, such as Yukawa or scalar self-couplings. We will study an example of this in the next section.

\section{Comparison to Other Coupling Variations} \label{sec:4dother}

In contrast to gauge couplings, other types of interactions, such as Yukawas or scalar self-couplings, are less constrained in their form. We demonstrate in this section that such couplings can be modified at tree-level by background scalar variations with a power-law dependence. Next, we apply this result to show that such couplings can induce in gauge couplings a power-law dependence on background scalars starting at two-loop order.

\subsection{Power-Law Variations in Yukawa Couplings}

Unlike gauge couplings, Yukawas and quartic couplings are not protected by gauge invariance from tree-level modifications by scalar backgrounds and can exhibit power-law sensitivity in four dimensions. To show the explicit difference, we construct two toy examples in the style of Froggatt-Nielsen~\cite{Froggatt:1978nt,Leurer:1992wg,Leurer:1993gy} below.

\subsubsection{A Simple Model for a Dynamical Yukawa} \label{sec:4dotherlin}

Consider the following left-chiral Weyl fermions: $\eta$, $\xi^c$, $\psi$, $\psi^c$. Fields without (with) a $c$ superscript transform with charge $1$ ($-1$) under a global U(1) symmetry. Furthermore, we introduce an SU($N$) gauge symmetry under which $\xi^c$ transforms as an anti-fundamental while all other fermion fields are singlets. The scalar sector of the theory contains an SU($N$) fundamental $\Phi$ and a singlet $\phi$, which are both U(1) singlets. The fields and their charges are summarized in Table~\ref{tab:powerlawyuk}. We are interested in constructing a scenario where the strength of the effective Yukawa coupling $\Phi\xi^c\eta$ depends in a power-law--like way on the value of the singlet field $\phi$.
\begin{table}[t!]
\centering
\renewcommand{\arraystretch}{1.3}
\begin{tabular}{|c|*{6}{p{0.5cm}|}}
\hline
 & \centering$\eta$ & \centering$\xi^c$ & \centering$\psi$ & \centering$\psi^c$ & \centering$\Phi$ & \centering$\phi$ \tabularnewline \hline
$SU(N)$ & \centering$\mathbbm{1}$ & \centering$\bar{\bm{N}}$ & \centering$\mathbbm{1}$ & \centering$\mathbbm{1}$ & \centering${\bm{N}}$ & \centering$\mathbbm{1}$ \tabularnewline \hline
$U(1)$  & \centering$1$ & \centering$-1$ & \centering$1$ & \centering$-1$ & \centering$0$ & \centering$0$ \tabularnewline \hline
\end{tabular}
\caption{Charges in a model with a dynamical Yukawa coupling of $\Phi\xi^c\eta$.}
\label{tab:powerlawyuk}
\end{table}

Given the charge assignments above, the Lagrangian contains the following mass terms and Yukawa interactions,
\begin{equation}
\begin{aligned}
-\lag &\supset \Phi\xi^c\left(y\eta+Y\psi\right)+ \left(m+f\phi\right)\psi^c\eta + \left(M+F\phi\right)\psi^c\psi + {\rm h.c.}\, ,
\end{aligned}
\end{equation}
where $y$, $Y$, $f$, and $F$ are constants. We will assume the masses satisfy $m\ll M$. In the absence of scalar VEVs, $\psi^c$ pairs up with $(m\eta+M\psi)/\sqrt{m^2+M^2}\simeq \psi$ to form a Dirac fermion of mass $\sqrt{m^2+M^2}\simeq M$ while $(M\eta-m\psi)/\sqrt{m^2+M^2}\simeq \eta$ and $\xi^c$ are massless Weyl fermions. With this hierarchy of masses, when integrating out the $\psi^{(c)}$ fields, the dimension-5 operator
\begin{equation}
\begin{aligned}
-\lag_{\rm eff} &\supset \frac{Yf\phi}{M}\Phi\xi^c\eta+ {\rm h.c.}
\end{aligned}
\end{equation}
is generated. At dimension-6, an operator containing two $\phi$ insertions is generated,
\begin{equation}
\begin{aligned}
-\lag_{\rm eff} &\supset -\frac{YfF\phi^2}{M^2}\Phi\xi^c\eta+ {\rm h.c.}\, ,
\end{aligned}
\end{equation}
and so on. Resumming these $\phi$ insertions leads to a compact form for the Yukawa coupling of $\Phi$ to $\eta$ and $\xi^c$,
\begin{equation}
\begin{aligned}
-\lag_{\rm eff} &\supset \left[y+\frac{Yf\phi}{M}\left(1-\frac{F\phi}{M}+\dots\right)\right]\Phi\xi^c\eta+ {\rm h.c.}=\left(y+\frac{Yf\phi}{M+F\phi}\right)\Phi\xi^c\eta+ {\rm h.c.}
\end{aligned}
\end{equation}
The $\phi$-dependent effective Yukawa coupling can therefore be written as
\begin{equation}
\begin{aligned}
y(\phi)=y+\frac{Yf\phi}{M+F\phi}\, ,
\end{aligned}
\end{equation}
with $M+F\phi$ the $\phi$-dependent $\psi$ mass. This model is therefore an example where a Yukawa coupling can depend linearly on the background value of a scalar field that does not come from the leading term in an expansion of a logarithm. Of course, if the field variation is such that $F\phi\gtrsim M$, this effective theory breaks down and needs to be reorganized.

\subsubsection{Dynamical Yukawas with Higher Powers}

It is possible to extend this setup slightly to generate Yukawas that vary faster-than--linearly with the value of the singlet scalar field. We will sketch out such an extension that leads to $\phi^2$ scaling with generalizations to higher powers, more or less straightforward. To accomplish this, we keep the same gauge and global symmetries, light fermions, and scalars as before and add two pairs of gauge-singlet Weyl fermions $\psi_{1,2}$ and $\psi_{1,2}^c$ that have charges $1$ and $-1$ under the global U(1). In addition, we introduce a discrete $\mathbb{Z}_2$ under which $\psi_{2}$, $\psi_{2}^{c}$, and $\phi$ are odd while all other fields are even. We summarize these charge assignments in Table~\ref{tab:powerlawyuk2}.
\begin{table}[t!]
\centering
\renewcommand{\arraystretch}{1.3}
\begin{tabular}{|c|*{8}{p{0.5cm}|}}
\hline
 & \centering$\eta$ & \centering$\xi^c$ & \centering$\psi_1$ & \centering$\psi_1^c$ & \centering$\psi_2$ & \centering$\psi_2^c$ & \centering$\Phi$ & \centering$\phi$ \tabularnewline \hline
$SU(N)$ & \centering$\mathbbm{1}$ & \centering$\bar{\bm{N}}$ & \centering$\mathbbm{1}$ & \centering$\mathbbm{1}$ & \centering$\mathbbm{1}$ & \centering$\mathbbm{1}$ & \centering${\bm{N}}$ & \centering$\mathbbm{1}$ \tabularnewline \hline
$U(1)$ & \centering$1$ & \centering$-1$ & \centering$1$ & \centering$-1$ & \centering$1$ & \centering$-1$ & \centering$0$ & \centering$0$ \tabularnewline \hline
$\mathbb{Z}_2$ & \centering$+$ & \centering$+$ & \centering$+$ & \centering$+$ & \centering$-$ & \centering$-$ & \centering$+$ & \centering$-$ \tabularnewline \hline
\end{tabular}
\caption{Charges in a model that leads to the Yukawa coupling of $\Phi\xi^c\eta$ scaling as $\phi^2$.}
\label{tab:powerlawyuk2}
\end{table}

The allowed mass terms and Yukawa interactions in this model are
\begin{equation}
\begin{aligned}
-\lag &\supset \Phi\xi^c\left(y\eta+Y\psi_1\right)+m\psi_1^c\eta+f\phi\psi_2^c\eta + 
\begin{pmatrix}
\psi_1^c & \psi_2^c
\end{pmatrix}
\begin{pmatrix}
M_1 & F_{12}\phi \\
F_{21}\phi & M_2
\end{pmatrix}
\begin{pmatrix}
\psi_1 \\
\psi_2
\end{pmatrix}
+ {\rm h.c.}
\end{aligned}
\end{equation}
As in the previous model, the effective Lagrangian involving $\eta$ and $\xi^c$ comes from integrating out the $\psi_{1,2}^{(c)}$ fields (which form a pair of Dirac fermions). The leading correction to the $\Phi\xi^c\eta$ Yukawa occurs at dimension-6,
\begin{equation}
\begin{aligned}
-\lag_{\rm eff} &\supset \frac{YfF_{12}\phi^2}{M_1M_2}\Phi\xi^c\eta+ {\rm h.c.}\, ,
\end{aligned}
\end{equation}
and the next at dimension-8,
\begin{equation}
\begin{aligned}
-\lag_{\rm eff} &\supset \frac{YfF_{12}^2F_{21}\phi^4}{M_1^2M_2^2}\Phi\xi^c\eta+ {\rm h.c.}\, .
\end{aligned}
\end{equation}
Summing these up, we obtain
\begin{equation}
\begin{aligned} 
-\lag_{\rm eff} &\supset \left[y+\frac{YfF_{12}\phi^2}{M_1M_2}\left(1+\frac{F_{12}F_{21}\phi^2}{M_1M_2}+\dots\right)\right]\Phi\xi^c\eta+ {\rm h.c.}\\
&=\left(y+\frac{YfF_{12}\phi^2}{M_1M_2-F_{12}F_{21}\phi^2}\right)\Phi\xi^c\eta+ {\rm h.c.}\, ,
\end{aligned}
\end{equation}
so that the effective Yukawa coupling is
\begin{equation}
\begin{aligned}
y(\phi)=y+\frac{YfF_{12}\phi^2}{M_1M_2-F_{12}F_{21}\phi^2}\, ,
\end{aligned}
\end{equation}
noting that $M_1M_2-F_{12}F_{21}\phi^2$ is the $\phi$-dependent determinant of the $\psi_{1,2}$ mass matrix, i.e. the $\phi$-dependent product of their masses. 

It is straightforward to extend this class of models to include larger discrete symmetries, such that
\begin{equation}
\begin{aligned}
y(\phi)&\simeq Y\left[\frac{\phi}{\Lambda(\phi)}\right]^p\, ,
\end{aligned}
\end{equation}
with $p>2$ by including more heavy fermion fields where $\Lambda(\phi)$ is an effective scale set by ($\phi$-dependent) masses and couplings. 

These models show that faster-than-linear changes in a Yukawa coupling with respect to variations of a singlet scalar $\phi$ are possible, in at least some region of field space, without the variation arising from the expansion of a logarithm. A similar story can be constructed for scalar interactions. Related models of varying Yukawas have been explored in the context of baryogenesis and early-universe phase transitions in, e.g., Refs.~\cite{Berkooz:2004kx,Baldes:2016gaf,Baldes:2016rqn,vonHarling:2016vhf,Lillard:2018zts}. Lastly, we mention that large Yukawa variations could also be arranged in Nelson-Strassler setups~\cite{Nelson:2000sn}, where coupling to conformal sectors can allow for large RG running.

The examples shown in this section also highlight crucial differences between gauge and Yukawa coupling variations. Gauge invariance and renormalizability strongly limit the interactions of gauge bosons and only allow dynamical changes from a scalar background at loop order, with the leading term logarithmic. In contrast, Yukawas and quartics face no such restriction and can vary at tree-level through mixing with a power law dependence on the scalar VEV whose exponent is controlled by symmetry.

\subsection{A Dynamical Gauge Coupling from Yukawa Back-Reaction}

The general arguments of Section~\ref{sec:generalscale} demonstrated that gauge couplings are logarithmically sensitive to scalar backgrounds at leading perturbative order in four-dimensional renormalizable theories. This restriction to one-loop order is, however, essential to the conclusion. At two-loop order and beyond, Yukawa and scalar self-couplings enter the gauge beta function and open a qualitatively new channel for scalar-background sensitivity.

To illustrate this effect, consider a gauge theory containing (light) charged fermions that have a Yukawa interaction with the coupling $y$. At two-loop order, this coupling then enters the RG evolution of the gauge coupling~\cite{Machacek:1983tz,Machacek:1983fi,Machacek:1984zw,Arason:1991ic}:
\beq
\frac{dg}{dt} \ \simeq \ - \frac{b_1}{(4\pi)^2}g^3 -\frac{b_2^g}{(4\pi)^4}g^5
- \frac{b_2^y}{(4\pi)^4}g^3y^2 \\
\ \simeq \ -\frac{b_1}{(4\pi)^2}\left[1+\frac{b_2^y}{b_1(4\pi)^2}y^2\right]g^3 \ ,
\nnmb
\eeq
where in the second line we have dropped the subleading $g^5$ term. Neglecting as well the running of the Yukawa $y$, which is formally higher-order, this expression integrates to
\beq
\frac{1}{g^2(\mu)} = \frac{1}{g^2(\Lambda)} + \frac{2b_1}{(4\pi)^2}\left[1+\frac{b_2^y}{b_1(4\pi)^2}y^2\right]\ln\!\lrf{\mu}{\Lambda} \ .
\eeq
At this level of approximation, the Yukawa coupling can be seen as a modification to the effective one-loop beta-function coefficient.

Now suppose the Yukawa coupling has a power-law dependence on a background scalar $\phi$ as discussed above: $y(\phi)\simeq Y[\phi/\Lambda(\phi)]^{p}$. The two-loop contribution to $1/g^2$ scales then as  $[\phi/\Lambda(\phi)]^{2p}$. For large field excursions, this induces a temporary power-law behavior in the gauge coupling. Although this back-reaction mechanism is parametrically sub-leading, it provides a qualitatively distinct source of gauge coupling variation that is technically natural within four-dimensional perturbative field theory. It is most significant when the Yukawa coupling is large, and the scalar field excursion is substantial, precisely the regime where the leading logarithmic contribution is also large. We note as well that this mechanism relies on retaining some additional light charged state with a varying Yukawa, in contrast to the threshold scenarios of Sec.~\ref{sec:4d} where the variation arises from the dynamics of potentially heavy fields.

\section{Dynamical Gauge Couplings Beyond Four Dimensions} \label{sec:xd}

Having focused so far on renormalizable UV completions that lead to non-standard gauge coupling variations, we now turn to more general completions. A natural setting for this is theories with extra spacetime dimensions, where the fundamental gauge coupling has a negative mass dimension. In this section, we present examples of gauge coupling variations in the context of extra dimensions. We also demonstrate how these realizations can motivate strong logarithmic variations and potentially also yield non-logarithmic evolution of gauge couplings.

\subsection{Charged Matter in a Flat Extra Dimension}
\label{sec:bulkcharged}

Recall that we found previously that to obtain a significant gauge coupling variation from charged matter, many matter multiplets are needed. A natural origin for a large multiplicity of charged states is the Kaluza-Klein~(KK) excitations of a bulk matter field. To illustrate this, we consider a minimal scenario in a five-dimensional flat space-time where one dimension is compactified on an interval $z\in [0, R]$. The bulk field content consists of an abelian gauge field, a Dirac fermion $\Psi$ with charge $Q$, and a real singlet scalar $\Phi$. The five-dimensional Lagrangian density is
\beq
\lag_5 = -\frac{1}{4}F_{MN}F^{MN} + \frac{1}{2}\bar{\Psi}i\Gamma^MD_M\Psi - M\bar{\Psi}\Psi - \frac{y}{\sqrt{R}}\,\Phi\bar{\Psi}\Psi + \frac{1}{2}|{\del\Phi}|^2- V(\Phi) \ .
\label{eq:5dlag}
\eeq
Taking Neumann boundary conditions for the scalar, the potential can induce a uniform background expectation value $\Phi = \sqrt{R}\,\phi$. For the fermion we impose mixed boundary conditions with $\left.\Psi_L\right|_0=0$ and $\left.\Psi_R\right|_R=0$~(or $(-,+)$ parities in an orbifold realization). These conditions forbid a zero mode and produce a tower of Dirac fermion KK modes with masses~\cite{Csaki:2004ay,Ponton:2012bi}
\beq
m_n^2(\phi) = (M+y\,\phi)^2 + \lrf{w_n}{R}^{\!2} \ ,
\eeq
where $n = 1,2,\ldots \in \mathbb{N}$ and the roots $w_n$ satisfy $\tan(w_n) = -w_n/(|M+y\phi|R)$. For $n\gg |M+y\phi|R$ their asymptotic values are $w_n\simeq (2n-1)\pi/2$.

The field-dependent KK masses of the charged bulk fermion impact the running of the effective four-dimensional gauge coupling. Relative to a low reference scale $\mu_0 < m_1$, summing over KK modes gives the running coupling
\beq
\frac{1}{g^2(\mu,\phi)} = \frac{1}{g^2(\mu_0,\phi)}
+ \tilde{b}_\Psi\sum_{n=1}^\infty\ln\!\lrf{\mu^2}{m_n^2}\Theta(\mu-m_n) \, ,
\label{eq:kkrun1a}
\eeq
where $\tilde{b}_\Psi= -4Q^2/(3(4\pi)^2)$ is the contribution to the running from each Dirac KK mode. Note that while we have formally extended the KK sum to infinity, only a finite number of modes contribute for any given scale $\mu$. Thus, we expect the expression above to be reliable provided the running scale is less than the cutoff of the theory, $\mu < \Lambda \simeq 4\pi^2/(g^2R)$~\cite{Chacko:1999hg,Ponton:2012bi}, corresponding to $n\lesssim 4\pi/g^2$ modes (for $|M+y\phi|\ll \Lambda$). It is instructive to consider Eq.~\eqref{eq:kkrun1a} in the limit of $\mu \gg \pi/R,\,|M+y\phi|$. The number of modes contributing to the sum is then $N\simeq \big[\mu R/\pi\big]\gg 1$, and we find (generalizing Refs.~\cite{Dienes:2011ja,Dienes:2011sa})
\beq
\frac{1}{g^2(\mu,\phi)} &=& \frac{1}{g^2(\mu_0,\phi)}
+ 2b_\psi\left[\ln\!\lrf{\mu R}{N\pi}+\ln\!\lrf{N^N}{N!}-\sum_{n=1}^N\ln\!\lrf{m_n R}{n\pi}\right]
\label{eq:kkrun1b}\\
&\simeq&
\frac{1}{g^2(\mu_0,\phi)}
- 2b_\psi\!\lrf{\mu R}{\pi} - b_\psi\sum_{n=1}^N\ln\lrf{m_n^2R^2}{n^2\pi^2} \, .
\nnmb
\eeq
Going up in energy, the number of charged KK states contributing to the gauge running accumulates, leading to a sum of logarithmic contributions that produces a power law dependence on the RG scale $\mu$. This power law is precisely what is expected for the running of the five-dimensional gauge coupling~\cite{Dienes:2011ja,Dienes:2011sa}.

We can use the expression of Eq.~\eqref{eq:kkrun1b} to obtain the dependence of the running gauge coupling on the singlet VEV $\phi$ if we assume, as before, that this VEV does not impact the UV gauge coupling. The key feature is that the fractional changes in the KK masses due to variations in $\phi$ become increasingly smaller at higher KK levels, and the sum of these changes is convergent. Thus, matching gauge couplings for different $\phi$ values at a sufficiently high running scale gives the approximate IR relation
\beq
\frac{1}{g^2(\mu_0,\phi)}
\simeq \frac{1}{g^2(\mu_0,\phivev)}
- \tilde{b}_\Psi\sum_{n=1}^\infty\ln\!\lrf{m_n^2(\phi)}{m_n^2(\phivev)} \ ,
\label{eq:kkrun2}
\eeq
where $\phivev$ is a reference VEV. While we have allowed the sum on modes above to extend to infinity, only KK modes with $n\pi/R\lesssim |M+y\phi|$ contribute in a significant way, and we expect this expression to provide a good estimate for $|M+y\phi|\ll \Lambda$. By combining Eqs.~(\ref{eq:kkrun1b}) and~(\ref{eq:kkrun2}), we obtain the running coupling for any scale $\mu$ and background $\phi$ within the region of validity of the KK theory.  

Our main conclusion from this construction is that the presence of a tower of KK modes enhances the cumulative effect of the $\phi$-dependent threshold corrections on the gauge coupling, but does not significantly alter their logarithmic functional form. While the multiplicity of modes increases with energy, the dependence of the gauge coupling on the scalar VEV remains controlled since only modes with number $n \lesssim |M+y\phi|R$ have a meaningful effect. This contrasts with the overall running of the gauge coupling with $\mu$, for which the number of modes contributing continues to increase indefinitely with energy such that the net contribution scales as a power law.

\subsection{Dynamical Radius in a Flat Extra Dimension}

In the scenarios we have studied to this point, the gauge coupling in the UV has been independent of the varying scalar VEV. A potential counterexample is an extra-dimensional gauge theory in which the volume of the compactified extra dimensions depends on the expectation value of a scalar field. 

To illustrate the relation, consider again a bulk gauge theory in a flat five-dimensional spacetime with the extra dimension compactified on an interval of length $R$ (or a circle of radius $R/2\pi$). The five-dimensional gauge theory is non-renormalizable and must be seen as an effective field theory with a cutoff $\Lambda$. Below the cutoff, the theory can be expanded in KK modes and the massless zero mode identified with a four-dimensional gauge boson. The leading-order relation between the corresponding four-dimensional gauge coupling in the low-energy effective theory and the underlying five-dimensional gauge coupling is
\beq
\frac{1}{g^2(\mu)} = \frac{R}{g_5^2(\mu)} \ .
\label{eq:5dmatch}
\eeq
This matching can be applied at any scale between $\mu \gtrsim \pi R^{-1}$, corresponding to the first KK mass, and the UV cutoff of the theory $\mu \lesssim \Lambda \sim (4\pi)^2/(g^2 R)$~\cite{Chacko:1999hg,Ponton:2012bi}, since the running agrees in both the KK and five-dimensional approaches~\cite{Dienes:2011ja,Dienes:2011sa}. We also note that the connection between the four- and five-dimensional couplings can be modified further by brane-localized vector boson kinetic terms~\cite{Ponton:2001hq,Carena:2002me,Carena:2002dz}. Combining Eq.~\eqref{eq:5dmatch} evaluated at $\mu \to \Lambda$ with the running from Eq.~\eqref{eq:kkrun1a} assuming a single charged bulk fermion $\Psi$ with negligible bulk mass, the four-dimensional gauge coupling at the low-energy scale $\mu_0 < m_1$ is
\beq
\frac{1}{g^2(\mu_0,R)} &=&
\frac{R}{g_5^2(\Lambda)} - \tilde{b}_\Psi\sum_{n=1}^{\infty}\ln\!\lrf{\Lambda^2}{m_n^2}\Theta(\Lambda-m_n)
\label{eq:5dmatchlo}\\
&\simeq&
\frac{R}{g_5^2(\Lambda)}
+ \frac{2\tilde{b}_\Psi\Lambda\,R}{\pi} \, .
\nnmb
\eeq
This expression connects the low-scale coupling with the high-scale five-dimensional value to one-loop order.

Now, suppose the radius of the extra dimension depends on the VEV of a singlet scalar, $R=R(\phi)$, and that the five-dimensional gauge coupling is set by UV dynamics that are independent of the low-energy compactification.  It then follows from Eq.~\eqref{eq:5dmatchlo} that the four-dimensional gauge coupling varies linearly with the radius. Depending on how the radius depends on the scalar VEV, the gauge variation with $\phi$ is potentially much more drastic than the logarithmic dependence found in all our previous constructions. The specific form of $R(\phi)$ relies in turn on the mechanism of radius stabilization. We do not address this here, but we study it below in the context of a warped extra dimension.

While a varying radius may allow for a faster variation of the gauge coupling than other mechanisms, radius changes in the present construction have significant further implications. In flat extra dimensions, as we are considering here, graviton zero modes exhibit universal behavior, implying that the effective Planck scale is also highly sensitive to the radius. Specifically, we have~\cite{Csaki:2004ay,Sundrum:2005jf},
\begin{align}
    \mpl^2 = 2 R M_*^3 \ ,    
\end{align}
where $\mpl$ is the effective four-dimensional Planck scale and $M_*$ is the scale of gravity in the five-dimensional theory. The cost of a varying gauge coupling in this realization is therefore a change in the strength of gravity at long distances, which can have very significant phenomenological and cosmological implications~\cite{Uzan:2010pm,Uzan:2024ded}. Furthermore, the low-energy properties of other fields that propagate in the bulk may also be altered by changes in $R$.

More complicated realizations of gauge theories in extra dimensions may allow for gauge variations with a fixed four-dimensional Planck scale. For example, consider a set of $n$ compactified flat extra dimensions with volume $\text{Vol}_n$ together with a gauge theory confined to a subspace of dimension $n'\leq n$ with volume $\text{Vol}_{n^\prime}$. The four-dimensional gauge coupling is given by
\beq
\frac{1}{g^2} \ \sim \ \frac{\text{Vol}_{n^\prime}}{g_{n^\prime}^2} \ ,
\eeq
while the effective Planck mass is
\beq
\mpl^2 \ \sim \ \text{Vol}_nM_n^{2+n} \ ,
\eeq
where $g_{n^\prime}$ is the $n^\prime$-dimensional gauge coupling and $M_n$ is the $n$-dimensional scale of gravity. Variations in $\text{Vol}_{n^\prime}$ that leave $\text{Vol}_n$ invariant would then allow for changes in the gauge coupling without impacting the effective Planck mass.

Many of the features of the simple extra-dimensional models presented here are realized in more complete approaches such as string theory and its extensions. The low-energy values of both the gauge couplings and the Planck mass generically depend on the expectation values of scalar fields such as the dilaton~\cite{Witten:1984dg,Witten:1985xb,Damour:1994zq} and moduli~\cite{Witten:1985xb,Kaplunovsky:1987rp,Dixon:1990pc} connected to the compactification to four dimensions. However, the specific relative dependence on the dilaton and moduli depends on how the gauge theory is realized~\cite{Uzan:2010pm,Uzan:2024ded}, such as from closed strings in heterotic models~\cite{Witten:1985xb}, or from open strings connected to branes~\cite{Blumenhagen:2006ci,Heckman:2010bq,Weigand:2010wm}. The very broad latter class offers a wider set of possibilities, including realizations with localized gauge fields where the fluctuations of a sub-volume might allow for gauge variations with a constant low-energy Planck scale.

\subsection{Dynamical Radius in a Warped Extra Dimension}

It is instructive to compare the results above for a flat extra dimension to a realization in five dimensions with warping. Specifically, we consider the Randall-Sundrum geometry of Refs.~\cite{Randall:1999ee,Csaki:2004ay,Sundrum:2005jf,Ponton:2012bi,Csaki:2015xpj} with constant curvature $k$ in the fifth dimension corresponding to a finite slice of AdS$_5$. The background metric is
\beq
\dd[]{s^2} &=& e^{-2ky} \eta_{\mu \nu} \dd[]{x}^{\mu} \dd[]{x}^{\nu} - \dd[]{y}^{2} 
=
\frac{1}{(kz)^2}\left(\eta_{\mu \nu} \dd[]{x}^{\mu} \dd[]{x}^{\nu} - \dd[]{z}^{2}\right) \, ,
\eeq
where $k$ is the constant curvature of the fifth dimension, $z = k^{-1}e^{ky}$, and the compact coordinate ranges are taken to be $y\in [0,R]$ or equivalently $z\in [k^{-1},L]$ with $L=k^{-1}e^{kR}$. The boundaries are identified with UV~($y=0$) and IR~($y=R$) branes. For this background, the effective four-dimensional Planck scale in terms of the fundamental five-dimensional gravity scale $M_*$ is~\cite{Randall:1999ee}
\beq
\mpl^2 = \frac{M_*^3}{k}\bigg(1-e^{-2kR}\bigg) 
= \frac{M_*^3}{k}\bigg(1-\frac{1}{k^2L^2}\bigg) \, .
\eeq
We focus on $k\sim M_*$ and $kL \sim \exp(kR) \sim 10^{15}$ such that the scenario can address the electroweak hierarchy~\cite{Randall:1999ee}.

Consider now a gauge theory in the bulk space with five-dimensional gauge coupling $g_5$. Expanding in Kaluza-Klein modes, there is a massless zero mode that can be identified with the gauge boson in the four-dimensional reduced theory. The effective gauge coupling $g$ is equal to~\cite{Davoudiasl:1999tf,Pomarol:1999ad,Chang:1999nh,Gherghetta:2000qt}
\beq
\frac{1}{g^2} = \frac{R}{g_5^2} = \frac{1}{g_5^2}\frac{\ln(kL)}{k} \ .
\label{eq:5dmatchrs}
\eeq
While this expression matches the flat space result of Eq.~\eqref{eq:5dmatch} when written in terms of the $y$-coordinate range defined above, its physical interpretation is not the same. The quantity $R = (y_{\rm max}-y_{\rm min})$ no longer coincides with the invariant volume of the extra dimension, and as a result, the dependence of the four-dimensional Planck scale on $R$ is very different from that of the gauge coupling. Crucially, we see that variations in the extent of the $z$ coordinate through $L$ can alter the gauge coupling while hardly impacting the effective Planck scale when $kL\gg 1$.

To investigate such potential gauge variations, we consider first the stabilization of the fifth dimension. With only gravity in the bulk, the energy of the system is invariant under changes in $L$. Fluctuations in the size of the fifth dimension thus correspond to a massless \emph{radion} scalar field. The Goldberger-Wise~(GW) mechanism~\cite{Goldberger:1999uk} addresses this problem by introducing a bulk scalar $\Phi$ with the potential~\cite{Goldberger:1999uk,Csaki:2015xpj}
\beq
V(\Phi) \ \supset \ \frac{1}{2}m^2\Phi^2 + \delta(z-z_{\rm UV})\,\lambda_{\rm UV}(\Phi^2-v_{\rm UV}^2)^2
+ \delta(z-z_{\rm IR})\,\lambda_{\rm IR}(\Phi^2-v_{\rm IR}^2)^2 \ .
\eeq
The first term is a bulk mass, while the second and third are brane-localized potentials. For large $\lambda,\,\lambda^\prime \gg 1$, the scalar bulk dynamics generates a potential for the radion that is minimized with~\cite{Goldberger:1999uk,Goldberger:1999un,Csaki:1999mp}  
\beq
kL 
\simeq 
\frac{\sqrt{24}\mpl}{\chi}
\simeq
\lrf{v_{\rm UV}}{v_{\rm IR}}^{\!1/\varepsilon} \, ,
\eeq
where $\chi$ is the canonically normalized radion field and $\varepsilon = -2+\sqrt{4+m^2/k^2}$~\cite{Goldberger:1999uk,Goldberger:1999un,Csaki:1999mp}. The electroweak hierarchy can be obtained with $1/\varepsilon \sim v_{\rm UV}/v_{\rm IR} \sim 10$. Small values of $\varepsilon$ correspond to $m^2/k^2 \ll 1$, in which case $\varepsilon \simeq m^2/4k^2$. 

Applying the GW result to Eq.~\eqref{eq:5dmatchrs} yields
\beq
\frac{1}{g^2} \ \simeq \  \frac{1}{g_5^2}\frac{\ln(\sqrt{24}\mpl/\chi)}{k}
\ \simeq \ \frac{1}{g_5^2}\frac{\ln(v_{\rm UV}/v_{\rm IR})}{\varepsilon k} \ \simeq \ \frac{1}{g_5^2}\frac{4k\ln(v_{\rm UV}/v_{\rm IR})}{m^2} \ .
\label{eq:5dmatchrsgw}
\eeq
This picture provides a concrete framework to investigate dynamical variations in the low-energy gauge coupling beyond the standard (four-dimensional) renormalization group evolution. We see immediately that the gauge coupling depends logarithmically on the expectation value of the canonically normalized radion $\chi$. In terms of the parameters of the GW model, and assuming that $g_5$ and $M_*$ are set by the UV structure and are not affected by the stabilization dynamics, the value of the gauge coupling $g$ can potentially vary with $k$, $m^2$, $V_{\rm UV}$, and $V_{\rm IR}$. Now, a key feature of the GW mechanism is that the backreaction of the bulk scalar on the bulk geometry is small provided $v_{\rm UV,IR}^2/M_*^3 \ll 1$~\cite{Goldberger:1999uk,Goldberger:1999un,Csaki:1999mp}, and in particular $k$ remains essentially unchanged in this limit. Variations in $m^2$ could arise by replacing the parameter with the VEV of another bulk scalar. However, introducing such a scalar is likely to modify the GW solution in a significant way, and we do not consider this possibility further. In contrast, the values of $v_{\rm UV}$ and $v_{\rm IR}$ could plausibly arise from the VEVs of brane-localized scalars while not impacting the GW solution. For example, consider modifying one or both of the brane potentials ($i= \text{UV,\,IR}$) to
\beq
V_i = \lambda_i\Phi^4 - \frac{1}{k}(\mu_i^2+\kappa_i\phi_i^2)\Phi^2+\Delta V_i(\phi_i) \ ,
\eeq
where $\phi_i$ is a four-dimensional field localized on the $i$-th brane and $\Delta V_i(\phi_i)$ is a potential to fix its VEV. This potential produces in the GW context
\beq
\Phi^2(z_i) \simeq
v_i^2 \equiv \frac{\mu_i^2+\kappa_i\phi_i^2}{2\lambda_i} \ .
\eeq
Inserting this form back into Eq.~\eqref{eq:5dmatchrsgw}, we find a logarithmic dependence of $g$ on $\phi_i$ as claimed.

Variations of $g$ with background singlet scalars in this five-dimensional RS picture can also be understood through holography in terms of a four-dimensional theory where the gauge boson interacts with an approximate conformal field theory~(CFT) as a weakly gauged subgroup of the global symmetries of the CFT~\cite{Arkani-Hamed:2000ijo,Rattazzi:2000hs}. Under the correspondence, the CFT is broken by irrelevant operators at scale $\mu \sim k$, and spontaneously (and weakly explicitly) by nearly marginal operators at scale $\mu \sim L^{-1}$, with $\mu \sim z^{-1}$ describing the running scale between these cutoffs. Returning to Eq.~\eqref{eq:5dmatchrs}, the expression gives $g(\mu=L^{-1})$, where now the logarithm describes the impact of the CFT on the running of the gauge coupling between $\mu \sim k$ (where there is a Landau pole) down to $\mu \sim L^{-1}$ with an effective beta-function coefficient due to CFT states of $\tilde{b}_{CFT} \simeq -1/(2g_5^2k)$~\cite{Pomarol:2000hp,Arkani-Hamed:2000ijo}. In this context, scalar dynamics that modify the RS bulk extent $L$ correspond to changing the mass threshold of the CFT matter. This is directly analogous to the simple charged matter portal discussed in Sec.~\ref{sec:4d}, but now with strongly coupled CFT states impacting the running instead of weakly-coupled, massive, charged fields.

\subsection{Dynamical Gauge Couplings in the Early Universe with Extra Dimensions}

\begin{figure}[t!]
    \centering
     \includegraphics[width=0.49\textwidth, scale=0.49]{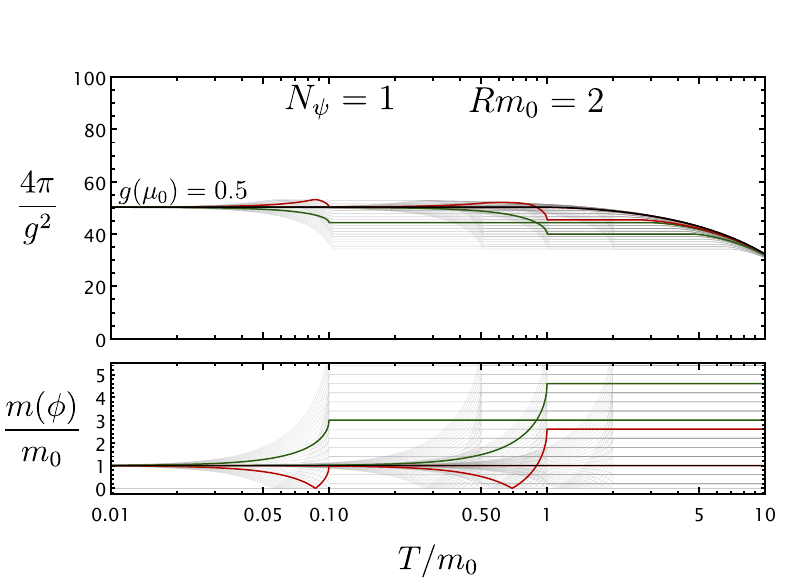}
 \includegraphics[width=0.49\textwidth, scale=0.49]{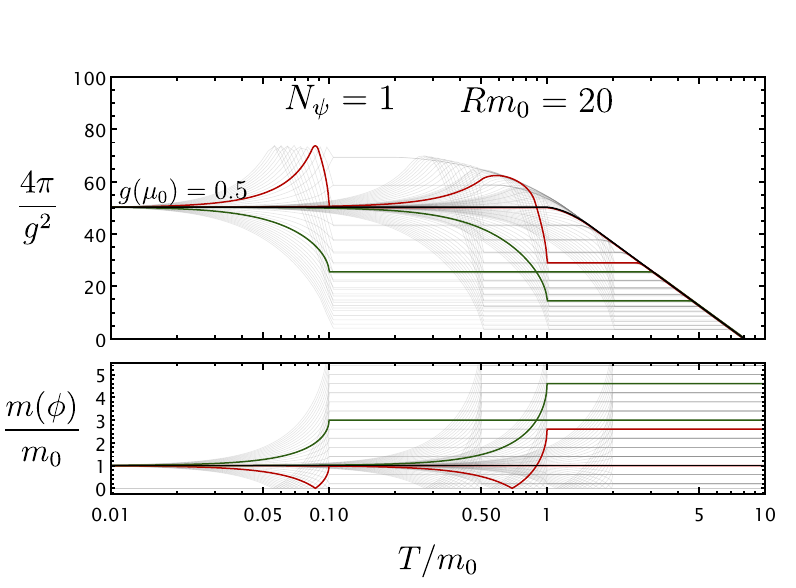}
    \caption{Inverse gauge coupling~(top panels) and bulk mass parameter~(bottom panels) as functions of temperature for the five-dimensional $U(1)$ toy model with a single bulk Dirac fermion, normalized to the fixed low-energy~(IR) coupling and mass $m_0\equiv m(\phi_0)$. The two columns correspond to small~(left) and large~(right) compactification radii. Red and green curves highlight specific trajectories with the same bulk mass parameter variation for the given critical temperature.}
    \label{fig:KK1}
\end{figure}

In this section, we present an explicit example of dynamical gauge variations in the early universe from a scenario with an extra dimension. As discussed above, there are two main ways for the gauge couplings to be affected: the masses of charged matter in the bulk can be shifted by a Higgs-like mechanism, thereby moving the thresholds of the charged KK tower; or the volume of compactified extra dimensions can change dynamically, which then impacts the effective four-dimensional gauge coupling derived from a gauge theory in the compactified bulk. Both approaches depend on the exact geometry of the extra-dimensional scenario, while their overall characteristics are independent of it. Here we focus on the first option, since the second depends in a complicated way on the compactification mechanism and its evolution with temperature.

The specific theory that we consider is the five-dimensional model of Sec.~\ref{sec:bulkcharged} with a flat extra dimension on an interval $z\in [0, R]$ containing a bulk $U(1)$ gauge theory, a bulk charged fermion $\Psi$, and a bulk singlet scalar $\Phi$ that couples to the fermion as described by Eq.~\eqref{eq:5dlag}. We assume that the bulk thermal potential for $\Phi$ generates a temperature-dependent expectation value $\Phi \to \sqrt{R}\,\phi$ with the temperature variation of $\phi$ given by Eq.~\eqref{eq:profile}. 

To illustrate the numerical impact of charged KK modes with varying masses on the effective gauge coupling, consider first the perspective of a low-energy observer with fixed low-scale gauge coupling and bulk mass $m_0 \equiv m(\phi_0) = |M+y\phi_0|$ in the background $\phi = \phi_0$ today. In Fig.~\ref{fig:KK1} we show evolution trajectories of the inverse gauge coupling with temperature normalized as $T/m_0$ for a single bulk fermion of unit charge in the top panel, and the fermion bulk mass $m(\phi) = |M+y\phi|$ in the bottom panel. Contours are shown for a range of values of $y\phi_0$ and $T_c/m_0 = [0.1,\,0.5,\,1.0,\,2.0]$, and we set $Rm_0 =2$~($Rm_0 = 20$) in the left~(right) panel. We also highlight two specific pairs of contours with green and red lines: the first pair has $T_c/m_0 = 0.1$ and $y\phi_0/m_0 = \pm 2.0$, while the second has $T_c/m_0 = 1.0$ and $y\phi_0/m_0 = \pm 4.0$.

In this theory, the change in the low-energy gauge coupling is sensitive to how many KK modes contribute at a given (low-energy) scale. The more modes stacked together, the faster the low-energy coupling changes (but also the closer it is to strong coupling). This feature can be seen by comparing the left and right panels in Fig.~\ref{fig:KK1}. For the smaller radius, $Rm_0=2$, the scenario is similar to the four-dimensional counterpart where only a single fermion contributes with a small coefficient. As the radius is made larger in the right panel, $Rm_0=20$, more modes can contribute in a small window of energy to produce a large effective number of fermions $\nff$. It should be noted that this region is close to the limit of where the KK description is reliable.

\begin{figure}[t!]
    \centering
     \includegraphics[width=0.49\textwidth, scale=0.49]{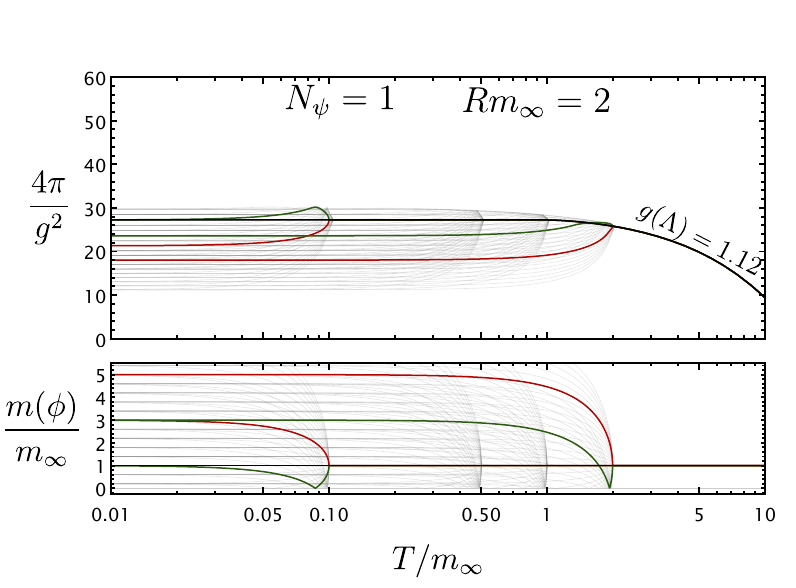}
 \includegraphics[width=0.49\textwidth, scale=0.49]{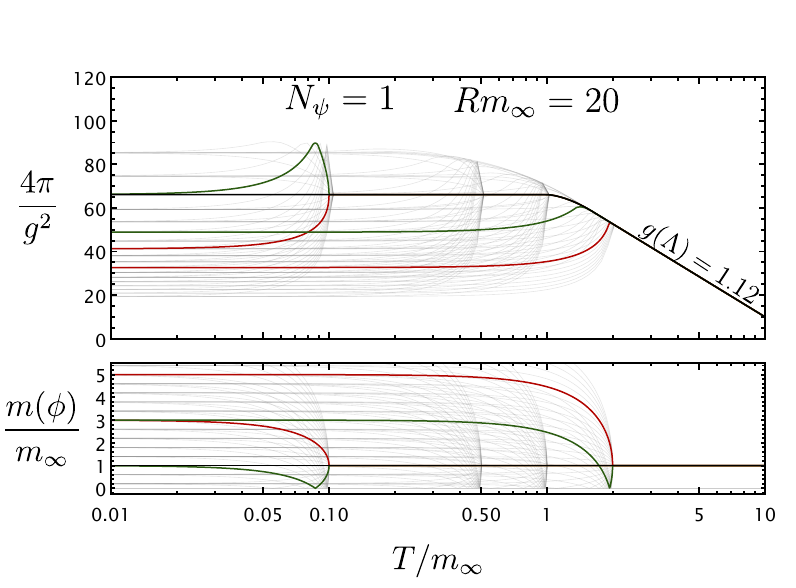}
    \caption{Same as Fig.~\ref{fig:KK1}, but with the gauge coupling normalized to a fixed high-scale (UV) value and all dimensionful scales normalized to the high-scale bulk mass parameter $m_\infty\equiv m(\phi=0)$.}
    \label{fig:KK2}
\end{figure}

Another instructive approach is to fix UV values of the gauge coupling and the bulk mass $m_\infty \equiv m(\phi=0) = M$ at some high scale, which we take to be $\mu = \Lambda = 5 m_\infty$. We show in Fig.~\ref{fig:KK2} trajectories of the inverse gauge coupling and the bulk mass parameter as functions of $T/m_\infty$ for a range of values of $y\phi_0$ and $T_c/m_\infty = [0.1,\,0.5,\,1.0,\,2.0]$, and we set $Rm_\infty =2$~($Rm_\infty = 20$) in the left~(right) panel. We also highlight two specific pairs of contours with green and red lines: the first pair has $T_c/m_\infty = 0.1$ and $y\phi_0/m_\infty = \pm 2.0$, while the second has $T_c/m_\infty = 1.0$ and $y\phi_0/m_\infty = \pm 4.0$. The trajectories shown in Fig.~\ref{fig:KK2} share many of the features of Fig.~\ref{fig:KK1}, and specifically a large coupling variation can arise with a single unit-charged bulk fermion for $Rm_\infty \gg 1$.

\section{Phenomenological Implications of Dynamical Gauge Couplings}\label{sec:pheno}

Our survey of mechanisms to alter the SM gauge couplings connects their dynamical changes to classes of new physics. In this section, we study direct bounds on such new physics and compare them with the extent to which the known gauge couplings can be altered, both in the early universe and today.

The modeling of dynamical gauge couplings explored in this work can be applied to any of the SM gauge groups. Among these, the fine-structure constant is the most precisely measured and most tightly constrained against cosmological variation, while the weak and strong couplings are comparatively less constrained before Big Bang nucleosynthesis (BBN) and more relevant for early-universe applications such as early weak and QCD confinement. 

\subsection{Implications of New Charged Matter}

A generic feature of all the renormalizable realizations of dynamical SM gauge couplings we have found is new matter charged under the SM gauge groups. To have avoided detection so far, this matter must typically be heavier than the weak scale and receive most of its mass independently of the Higgs, although see Ref.~\cite{Banta:2021dek} for a survey of potential exceptions. With renormalizability in mind as well, we focus here on new charged matter in the form of complex scalars or Dirac fermions. New charged matter faces very strong constraints if it is stable, which essentially rule it out assuming a standard cosmological history up to temperatures above the mass of the new particles~\cite{Perl:2009zz,Burdin:2014xma,DeLuca:2018mzn}. To avoid this, we focus on SM quantum numbers for the new charged particles that allow them to decay either directly to the SM or in a cascade that ends with a new neutral particle. 

For direct decays to the SM at the renormalizable level, the full set of possible SM quantum number assignments allowing this for a new Dirac fermion is collected in Ref.~\cite{Ishiwata:2015cga}, and a similar analysis can be done for the complex scalar. All the decay-allowing representations require that the new fields carry non-trivial electroweak quantum numbers. This implies that changes in one gauge coupling in this picture will typically be accompanied by changes in others. We also note that new interactions will generically lead to exotic flavor mixing~\cite{Ishiwata:2015cga}. However, there is a broad range of small cross-couplings that are consistent with current flavor measurements and allow decays with lifetimes below $\tau \lesssim 0.1\,\text{s}$ to avoid interfering with primordial nucleosynthesis~\cite{Kawasaki:2004qu,Jedamzik:2006xz}.

The second way to avoid charged relics is to have cascade decays of the charged states down to a lightest neutral species. A minimal realization is given by electroweak multiplet dark matter~\cite{Cirelli:2005uq}, but more possibilities exist. Cascade decays of new states carrying color charges also require that some of the new scalars or fermions carry electroweak charges. As for direct decays to the SM, this implies that variations in one of the SM gauge couplings would typically be correlated with others. Let us also point out that the neutral endpoint state of the cascade will contribute to the density of dark matter. For a standard hot early universe, its mass should not be much larger than the TeV scale to avoid generating too much DM~\cite{Griest:1989wd}.

New charged matter is also subject to direct search limits from colliders. These bounds are model-dependent, so we only give a broad summary here. Searches at the LHC for long-lived particles with lifetimes greater than $\tau\gtrsim 1\,\mu\text{s}$ impose the bounds $m\gtrsim 1300$--$2000\,\gev$ if they carry color charge, and $m\gtrsim 400$--$1200\,\gev$ if they are color-neutral but have unit electric charge~\cite{CMS:2024nhn,ATLAS:2025fdm}. An even wider spread in bounds is possible for new states that undergo cascade decays down to the lightest neutral particle. At the lower extreme, a unit-charged Higgsino-like fermion that is nearly degenerate with the neutral state is only limited to $m\gtrsim 140\,\gev$~\cite{ATLAS:2025lhc,CMS:2026ias}, while limits on cascade decays of strongly-interacting particles are typically above $m\gtrsim 1000\,\gev$~\cite{ATLAS:2021mdj}. As a working rule of thumb in what follows, we will impose the representative limits $m\gtrsim 100\,\gev$ for electrically charged particles and $m\gtrsim 1000\,\gev$ for new colour-charged states~\cite{Jeanty:2026etw}.

Similar considerations apply to extra-dimensional realizations of dynamical gauge couplings with bulk charged matter or some of the SM gauge fields propagating in the bulk.  Searches for dilepton and dijet resonances at the LHC push the first KK excitation of bulk graviton, electroweak, or color gauge bosons above a few TeV in mass,~\cite{ATLAS:2020fry,ATLAS:2025kmo,CMS:2026lsc,CMS:2026fjw}  with the precise bound depending on the localization of matter and the compactification geometry. This sets a lower bound on the compactification scale and correspondingly limits how much of the KK tower can contribute to coupling variations at cosmological temperatures near the weak scale or below.

\subsection{Implications and Dynamics of the Singlet Scalar}

Gauge variations in all the realizations we have considered have been controlled by the expectation value of a scalar field that is a singlet with respect to the low-energy gauge group. Changes in the scalar VEV can arise over the history of the cosmos as well as in specific media. These changes depend on how the scalar field couples to other matter, and are related to direct bounds on the scalar itself.  We study here the cosmological dynamics of a singlet scalar coupled to gauge-charged matter and its phenomenological implications. Our focus will be primarily on the charged matter realization of dynamical gauge couplings with a scalar $\phi$.

The cosmological evolution of the scalar $\phi$ depends crucially on how it connects with the surrounding thermal bath. In a sufficiently hot early universe, new charged matter thermalizes efficiently with its surroundings. Taking for concreteness $\nff$ flavors of fermionic charged matter $\psi$ with a Yukawa coupling to the scalar of the form of Eq.~\eqref{eq:yukawa}, reactions such as $\psi+\bar{\psi}\to \phi+A_\mu$ populate the scalar and drive it towards equilibrium with the rest of the plasma. These reactions are efficient when $T> m = \max\{m_\psi,m_\phi\}$ and we estimate that the scalar is fully thermalized by them for $\sqrt{\nff}y \gtrsim 10^{-7}(m/100\,\gev)^{1/2}$. The scalar could also be thermalized by other interactions it might have with the bath, such as a Higgs portal coupling~\cite{Piazza:2010ye}. Going forward, we will organize our analysis into two cases: i) a thermalized scalar with larger couplings to charged matter; ii) a non-thermalized scalar with very small couplings to charged matter.

For both cases, the coupling of the scalar to charged matter implies corrections to its effective potential at both zero and finite temperature. Consider charged fermion matter with $N_\psi$ fundamentals under the gauge group with mass $m(\phi) = |M+y\phi|$, as in Eq.~\eqref{eq:yukawa}. Without further structure in the theory~\cite{Brzeminski:2020uhm}, this implies unambiguous quantum corrections to the scalar potential with contributions to the zero-temperature scalar mass and quartic terms of size
\beq
\Delta m_\phi^2 \sim \frac{d(r_\psi)\nff\,y^2}{(4\pi)^2}M^2 \ ,\qquad
\Delta \lambda \sim \frac{d(r_\psi)\nff\,y^4}{(4\pi)^2} \ ,
\label{eq:natural}
\eeq
where $d(r_\psi)$ is the dimension of the fermion representation. Demanding minimal naturalness therefore puts lower bounds on these parameters. At finite temperature $T$, the charged matter is expected to thermalize efficiently, leading to a leading (one-loop) correction to the scalar potential for $T \gg m(\phi)$ of the form~\cite{Dolan:1973qd,Weinberg:1974hy,Anderson:1991zb}
\beq
\Delta V(\phi) \propto m_\psi^2(\phi)\,T^2 \ .
\label{eq:vt1loop}
\eeq
For $T\lesssim m(\phi)$ the one-loop contribution from charged matter falls off exponentially. However, there remains a two-loop correction from the induced coupling of $\phi$ to the vector boson of the form~\cite{Buchmuller:2003is,Cyncynates:2024bxw,Knapp-Perez:2025tns}
\beq
\Delta V(\phi) \propto g^2(\mu\sim T,\phi)\,T^4 \ .
\label{eq:vt2loop}
\eeq
We emphasize that this $T^4$ correction vanishes when $T \gtrsim m(\phi)$ in our charged matter model, and more generally in any scenario where the scalar background does not impact the (gauge) coupling above some mass threshold $m$ when $T > m$. Note as well that the scalar potential will receive additional corrections from any other couplings the scalar might have to light, thermalized matter, such as a Higgs portal interaction.

Turning now to the case of a thermalized scalar, the finite-temperature corrections to its potential can drive a phase transition in the early universe that changes $\phi$ and therefore the gauge coupling. This was the picture we considered previously in the charged matter model of Sec.~\ref{sec:4d}, where we found that a large dynamical variation of the gauge coupling due to the scalar $\phi$ requires a significant change in the mass of charged matter $m(\phi)$ at temperatures below the mass. Since the direct contribution of the charged matter to the effective potential is exponentially suppressed for $T < m(\phi)$, phase transitions relevant for dynamical gauge couplings must typically be driven by couplings of $\phi$ to thermalized fields that are lighter than the massive charged matter responsible for generating the field dependence of the effective gauge coupling. At the same time, the coupling of the scalar to charged matter, which is constrained to be heavier than the weak scale, together with naturalness, can make it challenging to push the critical temperature of a thermal phase transition involving $\phi$ much lower than the weak scale.

A further implication of a thermalized scalar is that the particle associated with it will obtain a significant population in the early universe. As the universe cools to below the mass of the physical excitation of the scalar around the background, which we write as $\phi(x) \to \phi +\eta(x)$, the density of $\eta$ particles will be depleted by decays. For the fermion coupling of Eq.~\eqref{eq:yukawa} and $\nff$ flavours with equal mass $m(\phi)$ in the gauge representation $r_\psi$, the leading decay channels are to fermions for $m_\eta > 2m$ and to vector bosons for $m_\eta < 2m$, with the decay rates
\beq
\Gamma(\eta\to\psi\bar{\psi}) &=& \frac{d(r_\psi)}{8\pi}\nff y^2m_\eta\left[1-\bigg(\frac{2m(\phi)}{m_\eta}\bigg)^2\right]^{3/2} \, ,
\label{eq:decvec1}\\
\Gamma(\eta\to AA) &=& \lrf{2S_2(r_\psi)\,\alpha}{3\pi}^{\!2}\!C_2(G)\,\nff^2y^2\,\frac{m_\eta^3}{m^2(\phi)}  \, ,
\label{eq:decvec2}
\eeq
where $d(r_\psi)$ is the dimension of the fermion representation, $S_2(r_\psi)$ is the trace invariant, and $C_2(G)$ is the group Casimir. These decays are relatively fast on cosmological timescales when the scalar thermalizes as long as the ratio $m_\eta/m(\phi)$ is not extremely small, and can easily eliminate the scalar density before primordial nucleosynthesis. Additional decay channels may open when the scalar has couplings to other light matter. We also note that for larger couplings the scalar can lead to diboson signals at colliders~\cite{Danielsson:2016nyy,Danielsson:2019ftq}.

The second case we consider is that of a non-thermalized scalar with very small couplings to charged or other matter. The time evolution of the scalar is now very different, and at late times in many scenarios is a coherent oscillation of the field rather than a thermal ensemble of particles. To see this, consider the limit of zero coupling ($y\to 0$) and let us approximate the scalar potential by $V(\phi)\simeq m_\phi^2\phi^2/2$, with $\phi=0$ being the minimum today. If the scalar is initially displaced from the origin by $\phi_i$, it remains trapped there by Hubble friction until $H \sim m_\phi$, after which the scalar rolls towards the minimum and begins to oscillate with frequency $m_\phi$ and amplitude $\bar{\phi}(t) = {\phi}_i(a_i/a)^{3/2}$. The energy density of the oscillations then redshifts like matter and contributes to the dark matter density today by the amount~\cite{Antypas:2022asj}
\beq
\fdm \ \simeq \
\bigg(\frac{m_\phi}{4.5\times 10^{-17}\,\text{eV}}\bigg)^{\!2}\bigg(\frac{\bar{\phi}(t_0)}{100\,\gev}\bigg)^{\!2} \, ,
\label{eq:uldm}
\eeq
where $\fdm = \rho_\phi/\rho_{DM}$ and $\bar{\phi}(t_0)$ is the amplitude today. Large field excursions therefore require an extremely small scalar mass to avoid producing too much dark matter. When $m_\phi \lesssim 10^{-20}\,\text{eV}$, ultralight scalars can interfere with standard cosmic structure formation and (for even lighter masses) alter the fractions of dark matter and dark energy from matter-radiation equality to today leading to stronger limits $f_{DM}\lesssim 10^{-2}$--$10^{-1}$~\cite{Antypas:2022asj}. Ultralight scalars can also be created through black hole superradiance, which damps the spin of the host black hole; this constrains decoupled ultralight scalars over the mass window $m_\phi \sim 10^{-13}$--$10^{-11}\,\text{eV}$~\cite{Arvanitaki:2014wva,Cardoso:2018tly}.

Returning to a non-zero coupling of the scalar to charged matter or other fields, this picture of an oscillating scalar can persist for sufficiently small couplings but is modified in several ways as the coupling increases. While the scalar itself need not be thermalized, its coupling to matter that is thermalized leads to thermal corrections to the scalar effective potential of the form of Eqs.~(\ref{eq:vt1loop},\,\ref{eq:vt2loop}). A specific example considered in Refs.~\cite{Batell:2021ofv,Chun:2021uwr} is an interaction with a fermion of mass $m(\phi)=|M+y\phi|$. Using this form in Eq.~\eqref{eq:vt1loop} generates a linear term in the scalar potential that can drive the scalar away from the origin while $T > m(\phi)$. When $T < m(\phi)$, this correction is suppressed, but there remains a two-loop contribution from the induced coupling of $\phi$ to the vector boson of Eq.~\eqref{eq:vt2loop}~\cite{Buchmuller:2003is,Cyncynates:2024bxw,Knapp-Perez:2025tns}. The net impact of thermal corrections to the scalar potential is typically to set an effective initial displacement $\phi_i$ of the scalar when its evolution transitions to oscillations in an $m_\phi^2\phi^2/2$ potential. 

Once scalar oscillations have begun, they can be damped by perturbative decays of the scalar~\cite{Dolgov:1982th,Abbott:1982hn} as well as non-perturbative effects analogous to inflationary pre-heating~\cite{Kofman:1994rk,Kofman:1997yn,Allahverdi:2010xz}. Damping by decays is negligible for small couplings and light scalar masses. For a scalar coupled to massive fermions with mass $m(\phi) = |M+y\phi|$, significant damping from fermion creation can arise for large oscillations with $|y\phi| > |M|$ when $y^2\phi^2/m_\phi^2 \gg 1$, but this effect turns off quickly once $|y\phi| < |M|$~\cite{Greene:1998nh,Giudice:1999fb,Greene:2000ew,Peloso:2000hy}.

\subsection{Dynamical Gauge Variations from a Thermalized Scalar}\label{sec:thermalized}

Having delineated two natural limits for the cosmological evolution of the singlet scalar that drives dynamical changes in one or more gauge couplings, we turn now to study some of the implications of our explicit constructions in the case where the scalar is thermalized with the SM in the early universe. This picture coincides naturally with a variation in the scalar background from a thermal phase transition. To be concrete, we focus on the charged matter UV completion of a dynamical gauge coupling and consider its implications for the phenomenological applications of Refs.~\cite{Berger:2019yxb,Berger:2020maa,Howard:2021ohe,Ellis:2019flb,Dvali:1995ce,Choi:1996fs,Croon:2022gwq,Ipek:2018lhm,Lu:2022yuc,Lohitsiri:2019wpq,Bhalla-Ladd:2025agq,Heurtier:2021rko}, where the $SU(2)_L$ electroweak or $SU(3)_c$ strong gauge couplings are required to differ significantly from their expected SM values at temperatures near the weak scale.

The gauge changes studied in most of these works were parameterized in terms of linear effective operators of the form of Eq.~\eqref{eq:fdp}. Instead, we consider the charged matter UV completion of these operators consisting of a singlet scalar $\phi$ with value $\phi_0$ today that couples to new matter charged under $SU(3)_c$ or $SU(2)_L$ with mass $m(\phi)$. For scales $\mu < m(\phi)$ and small variations, the dynamical change in the gauge coupling relative to the running SM value $\alpha_i(\mu,\phi_0)=g_i^2/4\pi$ (extrapolated using only SM matter) is from Eq.~\eqref{eq:logmatter2}
\beq
\Delta\alpha_i(\mu,\phi) \equiv \alpha_i(\mu,\phi)-\alpha_i(\mu,\phi_0) 
\simeq 
\frac{b_{\Delta}}{2\pi}\alpha_i^2(\mu,\phi_0)\ln\!\lrf{m(\phi)}{m(\phi_0)}\ ,
\label{eq:delalfi}
\eeq
where $b_\Delta =(4\pi)^2\,\tilde{b}_\Delta$ is the change in the relevant one-loop beta function coefficient due to the heavy matter, equal to $b_\Delta = 2\nff/3$ for $\nff$ Dirac fermion fundamentals under $SU(N)$.

As a first application of this result, we relate it to the scenario studied in Ref.~\cite{Ellis:2019flb}, which investigated the impact of dynamical gauge couplings on baryogenesis at temperatures near the weak scale. In this work, it was demonstrated that variations $\Delta\alpha_2 > 0$ or $\Delta\alpha_3 < 0$ at the time of the electroweak phase transition $T\sim 100\,\gev$ could enable the creation of the observed baryon asymmetry through electroweak baryogenesis~\cite{Ellis:2019flb}. To generate enough baryons in the scenario with $\Delta\alpha_3=0$, a positive change of at least $\Delta\alpha_2 \gtrsim 0.01$ is needed. With the expression of Eq.~\eqref{eq:delalfi}, this translates into $b_\Delta\ln(m(\phi)/m(\phi_0)) \gtrsim 55$ or equivalently $\nff  \gtrsim 83$ Dirac fermion fundamentals for $\ln(m/m_0)=1$. For $\Delta\alpha_2=0$, sufficient baryon production is obtained for $\Delta\alpha_3 \lesssim -0.03$, which requires $b_\Delta\ln(m(\phi)/m(\phi_0)) \lesssim -16$ or equivalently $\nff  \gtrsim 24$ Dirac fermion fundamentals for $\ln(m/m_0)=-1$. With either option, we see that a large multiplicity of charged matter is needed to achieve the desired effect. Both imply that the corresponding gauge couplings evolve to strong coupling in the UV in the absence of any other new physics. While the onset of strong coupling lies beyond current and future sensitivities, for $m(\phi) \sim \tev$ it occurs below the Planck scale and therefore requires a drastic change in the UV structure relative to the SM. The logarithmic dependence on the singlet scalar implies further that the required multiplicity of charged matter can only be reduced moderately with very large field excursions.

Another application of a dynamical gauge coupling has been to raise the scale of $SU(3)_c$ confinement in the early universe, often above the weak scale. Such an early period of confinement can impact and enable weak-scale dark matter~\cite{Berger:2020maa,Howard:2021ohe}, baryogenesis~\cite{Ellis:2019flb}, and axions~\cite{Dvali:1995ce,Choi:1996fs,Heurtier:2021rko}. Most of this earlier work used the effective operator of Eq.~\eqref{eq:fdp}. Instead, applying the charged matter model and using Eq.~\eqref{eq:logmatter2} while neglecting quark mass thresholds (valid before electroweak symmetry breaking in the early universe), we obtain
\beq
\Lambda(\phi) \simeq \tilde{\Lambda}_{\rm QCD}\lrf{m(\phi)}{m(\phi_0)}^{\!{b_\Delta}/b_{QCD}} \, ,
\label{eq:lqcdeff}
\eeq
where $\tilde{\Lambda}_{\rm QCD}\simeq 50\,\mev$ is the approximate QCD confinement scale in the absence of quark masses, $b_{\rm QCD} = 7$, and $b_\Delta = (2/3)\nff$ if the charged matter consists of $\nff$ Dirac fermion fundamentals. It is straightforward but tedious to extend this result to include quark mass thresholds that lie above the modified confinement scale. Note that Eq.~\eqref{eq:lqcdeff} only holds for $m(\phi) > \Lambda(\phi)$; when this condition is not met, the effective confinement scale lies below the QCD value.  

For many of the applications of early QCD confinement envisaged in Refs.~\cite{Berger:2020maa,Howard:2021ohe}, the strong coupling scale $\Lambda(\phi)$ should be increased above the scale of electroweak symmetry breaking in the early universe, $\Lambda(\phi) \gtrsim 100\,\gev$. To achieve this, the enhancement factor in Eq.~\eqref{eq:lqcdeff} must be greater than $1000$, corresponding to $m(\phi)/m(\phi_0) \gtrsim 1400$ for $\nff = 10$ Dirac fermion fundamentals, $m(\phi)/m(\phi_0) \gtrsim 10$ for $\nff = 30$, and $m(\phi)/m(\phi_0) \gtrsim 2$ for $\nff = 100$. As in the previous example, with no other new physics, such a large multiplicity of charged matter drives the QCD coupling to large values in the UV, destroying asymptotic freedom. The new matter would also provide a spectacular target for future hadron collider searches.

\subsection{Dynamical Gauge Variations from an Unthermalized Scalar}

Dynamical gauge variations can also be generated by an unthermalized scalar. These provide a mechanism for gauge changes at temperatures well below the weak scale if they are ultralight. Specifically, starting from an initial displacement, they are expected to oscillate freely once the Hubble rate falls below their mass. In this context, we note that $m_\phi \simeq 10^{-28}\,\text{eV}$ for oscillations starting at recombination, $m_\phi \simeq 10^{-27}\,\text{eV}$ for matter-radiation equality, $m_\phi \simeq 10^{-15}\,\text{eV}$ for $T \simeq 1\,\mev$, and $m_\phi \simeq 10^{-5}\,\text{eV}$ for $T \simeq 100\,\gev$. This enables a significant variation in the mean gauge coupling from before to after the onset of oscillations, as well as an oscillating variation that can persist until today. However, the magnitudes of such variations are constrained by direct limits on new forces induced by the exchange of the physical excitation of the scalar field. We study this here with a focus on a charged completion that connects the scalar to the SM photon and gluon.

A standard parametrization of the leading coupling of the physical scalar $\eta(x)=\phi(x)-\phi$ to the photon and gluon (and light quarks) is~\cite{Kaplan:2000hh,Damour:2010rm,Damour:2010rp,Antypas:2022asj}
\begin{align}
\lag \ \supset \ \sum_{n\geq 1}\frac{1}{n}\!\lrf{\eta}{\sqrt{2}\mpl}^{\!n}\left[\frac{d_e^{(n)}}{4e^2}F_{\mu\nu}F^{\mu\nu}- \frac{\beta_3d_g^{(n)}}{2g_3^3}G_{\mu\nu}^aG^{a\,\mu\nu}
-\sum_{q}\big(d_{m_q}^{(n)}+\gamma_md_g^{(n)}\big)m_q\bar{q}q
\right] \, ,
\label{eq:dn}
\end{align}
where $\beta_3$ is the QCD beta function, and $\gamma_m$ is the anomalous dimension of the running quark masses. The normalization of the couplings to gluons and quarks is chosen to produce a combination of operators that is renormalization-group invariant~\cite{Kaplan:2000hh}. This form can be connected with the predictions of a given UV completion of the effective coupling of $\phi$ to the vector bosons.

To illustrate the matching for the $n=1$ operator, consider again the charged matter model with Dirac fermions of mass $m(\phi)$ transforming under the representation $(r_3, r_2, Y)$ of $SU(3)_c\times SU(2)\times U(1)_Y$. From Eq.~\eqref{eq:logmatter2}, we have the full one-loop dependence of the effective gauge coupling on the background $\phi$ at scale $\mu < m(\phi)$. Up to operators with more derivatives, the coupling of the $\eta$ excitation to the vector bosons is obtained by expanding $\phi\to \phi+\eta$~\cite{Shifman:1978zn}. This yields a linear coupling to the gluon in the present vacuum $\phi_0$ of
\beq
d_g^{(1)} = \left.\frac{b_{\Delta,3}}{2b_3}\frac{\sqrt{2}\mpl}{m}\frac{dm}{d\phi}\right|_{\phi=\phi_0} \ ,
\qquad
d_{m_q}^{(1)} = -\gamma_m d_g^{(1)} \ .
\label{eq:dg1}
\eeq
Here, $b_3 = 11 - 2N_f/3$ and $b_{\Delta,3} = (4/3)d(r_2)S_2(r_3)$.
Note that to leading order in the gauge interaction, there is no coupling of $\phi$ to the quark mass operator at the matching scale, so the second relation is needed to match the model with the parametrization of Eq.~\eqref{eq:dn}. 

For the leading photon coupling in the charged matter model, the matching scale $m(\phi)$ must typically lie at or above the weak scale given the direct bounds on new charged particles. The effective coupling to the photon after electroweak symmetry breaking is then
\beq
\frac{1}{e^2(\mu,\phi)}
=\frac{1}{e^2(\mu,\phi_0)} 
- 2\tilde{b}_{\Delta,e}
\ln\!\lrf{m(\phi)}{m(\phi_0)} \, ,
\eeq
where $\tilde{b}_{\Delta,e} = \tilde{b}_{\Delta,2}+\tilde{b}_{\Delta,Y}
= [4/3(4\pi)^2]d(r_3)\big[S_2(r_2)+d(r_2)Y^2\big]$. Note that we have assumed $m(\phi)> \mu$. Expanding and connecting with Eq.~\eqref{eq:dn}, we find
\beq
d_e^{(1)} = \left.\frac{\alpha\, b_{\Delta,e}}{2\pi}\frac{\sqrt{2}\mpl}{m}\frac{dm}{d\phi}\right|_{\phi=\phi_0} \, ,
\label{eq:de1}
\eeq
with $b_{\Delta,e}=(4\pi)^2\tilde{b}_{\Delta,e}$. 

The operator coefficients $d_i^{(1)}$ can be related to gauge coupling variations once the field dependence of the mass $m(\phi)$ is specified. To illustrate this, let us focus on the electromagnetic coupling and take the fermion mass to be $m(\phi)=M+y\phi$, for which $d_e^{(1)}=y\, \alpha\,b_{\Delta,e}\sqrt{2}\mpl/2\pi m(\phi_0)$. The leading change in $\alpha$ is then 
\beq
\frac{\Delta\alpha}{\alpha} \simeq \frac{d_e^{(1)}}{\zeta}\ln(1+\zeta\varphi) \ ,
\label{eq:alfvar1}
\eeq
where $\zeta = y\sqrt{2}\mpl/m(\phi_0)$ and $\varphi = (\phi-\phi_0)/\sqrt{2}\mpl$. For fixed $d_e^{(1)}$, the variation is approximately independent of $\zeta$ for $\zeta \lesssim 1/\varphi$ and is suppressed due to the logarithm otherwise. In the minimal picture of a freely oscillating scalar, $\varphi$ is displaced with its maximal amplitude at early times when $m_\phi < H$, after which time it oscillates freely with mean value zero and amplitude falling as $a^{-3/2}$. The initial amplitude of $\varphi$ is determined by the mass $m_\phi$ and the dark matter fraction $\fdm$ via Eq.~\eqref{eq:uldm} together with solving for the redshift at which $H\simeq 3m_\phi$; we find that it is always less than unity for $\fdm \leq 1$. The expression for the leading variation in $\alpha_3$ is identical but with $d_e^{(1)}$ replaced by $d_g^{(1)}$ and an additional overall factor of $b_3\alpha_3/\pi$ due to the different normalization in Eq.~\eqref{eq:dn}. 

Variations in $\alpha$ and $\alpha_3$ from an ultralight scalar are limited by direct bounds on the $d_i^{(1)}$ coefficients. For $m_\phi \lesssim 10^{-5}\,\text{eV}$, there are strong bounds on the linear couplings of Eqs.~(\ref{eq:dg1},\,\ref{eq:de1}) due to apparent violations of the equivalence principle from the exchange of the light scalar; bounds range between $d^{(1)}_{e,g}\lesssim 10^{-5}$--$10^{-1}$, depending on the mass of the scalar~\cite{Antypas:2022asj,Hees:2018fpg}. When $m_\phi \lesssim 10^{-18}\,\text{eV}$, the oscillation frequency of the electromagnetic coupling from a background density of the oscillating scalar can match the sensitivity ranges of precision atomic spectroscopy in Rb and Cs atoms, which are primarily sensitive to $\alpha$ and nuclear mass changes via $\alpha_3$~\cite{Hees:2016gop,Hees:2018fpg}. These measurements limit $d_e^{(1)}\lesssim 10^{-8}f_{\rm DM}^{1/2}$ (in isolation) for $m_\phi \simeq 10^{-23}\,\text{eV}$ and exceed constraints from fifth forces for $f_{\rm DM} \sim 1$ over a wide mass range around this. Recent measurements with Th-229 nuclear clocks  provide improved bounds up to $m_\phi \simeq 10^{-18}\,\text{eV}$~\cite{Arakawa:2026mls,DeCol:2026hnl}. Within our charged matter model, these limits often exclude coupling variations that are consistent with the naturalness conditions of Eq.~\eqref{eq:natural} without additional structure in the theory. 

Larger gauge variations are possible when the leading interaction is quadratic in the scalar field, $n=2$ in Eq.~\eqref{eq:dn}. As discussed in Sec.~\ref{sec:symmetry}, this can arise when $\phi$ is charged under a symmetry that is realized linearly at $\phi = \phi_0$. Taking as a minimal example a theory that is invariant under $\phi\to -\phi$, that has charged matter with mass $m^2(\phi) = M^2\mp y^2\phi^2$, and $\phi_0 = 0$ today, we find $d_i^{(1)}=0$ together with the quadratic coefficients
\beq
d_g^{(2)} = \mp \frac{b_{\Delta,3}}{b_3}\frac{2\mpl^2}{M^2}y^2 \ ,
\qquad
d_e^{(2)} = \mp \frac{\alpha b_{\Delta,e}}{2\pi}\frac{2\mpl^2}{M^2}y^2 \ .
\eeq
As before, we can express changes in the gauge couplings in terms of these coefficients. For the electromagnetic coupling in this example, we find for $m(\phi) > \mu$
\beq
\frac{\Delta \alpha}{\alpha} \simeq \mp\frac{d_e^{(2)}}{2\zeta^2}\ln(1\mp\zeta^2\varphi^2) \ ,
\eeq
with a similar expression for the strong coupling. As in Eq.~\eqref{eq:alfvar1} for the linear variation, we see that the logarithmic form becomes important for $|\zeta\varphi| \gtrsim 1$. For the negative sign, note that the argument of the logarithm is always positive over the region of validity of this expression; before it goes negative, we encounter $m(\phi) \leq \mu$ and the gauge change then becomes independent of $\phi$, as seen from Eq.~\eqref{eq:logmatter2}.

Direct limits on $d_i^{(2)}$ assuming vanishing $d_i^{(1)}$ are collected in Ref.~\cite{Hees:2018fpg}. They are much less stringent than those on linear couplings and allow for a greater variation in gauge couplings. Phenomenological implications of this freedom have been studied in Refs.~\cite{Olive:2007aj,Sibiryakov:2020eir,Banerjee:2022sqg,Bouley:2022eer,Becker:2025pgb,Ghosh:2025pbn,Gan:2025nlu,Bouley:2026frx,Brzeminski:2026pgz}.

\section{Conclusion}\label{sec:conc}

In this work, we have investigated mechanisms for dynamical gauge couplings, and we have studied the extent to which they could plausibly vary in the early universe. The mechanisms we have considered connect the gauge kinetic operator to a scalar field. If the scalar gets an expectation value, it will alter the effective gauge coupling at low energies. In weakly-coupled renormalizable theories in four dimensions, we find that the form of the operator and therefore the size of the gauge variation is tightly constrained by symmetry and RG structure, and is generically logarithmic in the scalar VEV. This has immediate consequences for model building, and scenarios that rely on large or rapid changes in gauge couplings are challenging to achieve. 

These limitations follow from the general structure of scale anomaly matching across massive thresholds and are not specific to any particular UV completion. We highlight two ways to avoid this limitation. First, beyond leading order, the running gauge couplings obtain a dependence on Yukawa couplings in the theory that connect to charged matter. Such Yukawa couplings can vary with power-law dependence on the scalar background, which then translates into faster-than-logarithmic dependence of the gauge coupling at two loops. And second, going beyond four dimensions, if the gauge field propagates in one or more compactified extra dimensions, the corresponding coupling becomes tied to the volume of those dimensions. Radion dynamics that alter this volume can then induce a power-law variation in the gauge coupling at the classical level. 

In contrast to gauge couplings, Yukawa couplings are not protected by gauge invariance from tree-level modifications from scalar backgrounds. As we demonstrated in Sec.~\ref{sec:4dother}, UV completions in the spirit of Froggatt–Nielsen~\cite{Froggatt:1978nt} can generate effective Yukawa couplings of the form $y(\phi)\simeq Y[\phi/\Lambda(\phi)]^p$. This power-law dependence is qualitatively different from the logarithmic behavior forced upon gauge couplings by gauge invariance and the structure of the one-loop effective action, and makes dynamical Yukawa scenarios more flexible than their gauge counterparts. 

Our results have significant implications for scenarios that rely on dynamical gauge couplings to address phenomenological puzzles. The logarithmic limitation on four-dimensional gauge coupling variations is a robust consequence of perturbative quantum field theory. In general, order-one changes in the gauge coupling require either large field excursions, many charged states, or large couplings near the edge of perturbativity. More generally, a careful treatment of the underlying effective field theory and its potential origin is essential when assessing such scenarios.

\acknowledgments

We thank Brian Batell, Seyda Ipek, Kristan Jensen, Gopolang Mohlabeng, and Tim Tait for helpful discussions. This work is supported by Discovery Grants from the Natural Sciences and Engineering Research Council of Canada~(NSERC). TRIUMF receives federal funding via a contribution agreement with the National Research Council~(NRC) of Canada. MS is supported by KIAS individual grant PG107401.


\bibliographystyle{JHEP}
\bibliography{bibchange.bib}

\end{document}